\documentclass[apj]{emulateapj}

\usepackage{graphicx}
\usepackage{apjfonts}
\usepackage{natbib}
\usepackage[colorlinks,citecolor=blue,linkcolor=blue,urlcolor=blue,dvips]{hyperref}

\newcommand{\erg}{${\rm erg}\ {\rm cm}^{-2}\ {\rm s}^{-1}$} 
\newcommand{\phcms}{${\rm photons}\ {\rm cm}^{-2}\ {\rm s}^{-1}$} 
\newcommand{\Fermi}{{\it Fermi}}

\newcommand{\Swift}{{\it Swift}}

\newcommand{\NuSTAR}{{\it NuSTAR}}
\newcommand{\XMM}{\textit{XMM-Newton}}
\newcommand{\degree}{$^{\circ}$}
\newcommand{\trise}{\tau_{\rm rise}}
\newcommand{\tfall}{\tau_{\rm fall}}

\newcommand {\eg} {{\it e.g.}}

\newcommand {\be} {\begin{equation}}
\newcommand {\ee} {\end{equation}}
\newcommand {\bea} {\begin{eqnarray}}
\newcommand {\eea} {\end{eqnarray}}

\slugcomment{accepted to ApJ}

\shorttitle{3C 279 flares in 2013$-$2014}
\shortauthors{Hayashida et~al.}

\begin{document}

\pagenumbering{arabic}


\title{Rapid Variability of Blazar 3C~279 during Flaring States in 2013$-$2014 with Joint 
\Fermi-LAT, \NuSTAR, \Swift, and Ground-Based Multi-wavelength Observations}


\author{
M.~Hayashida\altaffilmark{1,2}, 
K.~Nalewajko\altaffilmark{3,4,5}, 
G.~M.~Madejski\altaffilmark{3}, 
M.~Sikora\altaffilmark{6},
R.~Itoh\altaffilmark{7},
M.~Ajello\altaffilmark{8}, 
R.~D.~Blandford\altaffilmark{3}, 
S.~Buson\altaffilmark{9,10}, 
J.~Chiang\altaffilmark{3}, 
Y.~Fukazawa\altaffilmark{7},
A.~K.~Furniss\altaffilmark{3},
C.~M.~Urry\altaffilmark{11}, 
I.~Hasan\altaffilmark{11}, 
F.~A.~Harrison\altaffilmark{12}, 
D.~M.~Alexander\altaffilmark{13},
M.~Balokovi\'{c}\altaffilmark{12}, 
D.~Barret\altaffilmark{14,15},
S.~E.~Boggs\altaffilmark{16},
F.~E.~Christensen\altaffilmark{17},
W.~W.~Craig\altaffilmark{16,18},
K.~Forster\altaffilmark{12},
P.~Giommi\altaffilmark{19},
B.~Grefenstette\altaffilmark{12},
C.~Hailey\altaffilmark{20},
A.~Hornstrup\altaffilmark{17},
T.~Kitaguchi\altaffilmark{21},
J.~E. Koglin\altaffilmark{3},
K.~K.~Madsen\altaffilmark{12},
P.~H.~Mao\altaffilmark{12},
H.~Miyasaka\altaffilmark{12},
K.~Mori\altaffilmark{20},
M.~Perri\altaffilmark{19,22},
M.~J.~Pivovaroff\altaffilmark{18}, 
S.~Puccetti\altaffilmark{19,22},
V.~Rana\altaffilmark{12},
D.~Stern\altaffilmark{23},
G.~Tagliaferri\altaffilmark{24},
N.~J.~Westergaard\altaffilmark{17},
W.~W.~Zhang\altaffilmark{25},
A.~Zoglauer\altaffilmark{16},
M.~A.~Gurwell\altaffilmark{26},
M.~Uemura\altaffilmark{27},
H.~Akitaya\altaffilmark{27}, 
K.~S.~Kawabata\altaffilmark{27},
K.~Kawaguchi\altaffilmark{7},
Y.~Kanda\altaffilmark{7},
Y.~Moritani\altaffilmark{27}, 
K.~Takaki\altaffilmark{7},
T.~Ui\altaffilmark{7}, 
M.~Yoshida\altaffilmark{27},
A.~Agarwal\altaffilmark{28},
A.~C.~Gupta\altaffilmark{28}
}
\altaffiltext{1}{Institute for Cosmic Ray Research, University of Tokyo, 5-1-5 Kashiwanoha, Kashiwa, Chiba, 277-8582, Japan}
\altaffiltext{2}{email:mahaya@icrr.u-tokyo.ac.jp}
\altaffiltext{3}{W. W. Hansen Experimental Physics Laboratory, Kavli Institute for Particle Astrophysics and Cosmology, Department of Physics and SLAC National Accelerator Laboratory, Stanford University, Stanford, CA 94305, USA}
\altaffiltext{4}{NASA Einstein Postdoctoral Fellow}
\altaffiltext{5}{email:knalew@stanford.edu}
\altaffiltext{6}{Nicolaus Copernicus Astronomical Center, 00-716 Warsaw, Poland}
\altaffiltext{7}{Department of Physical Sciences, Hiroshima University, Higashi-Hiroshima, Hiroshima 739-8526, Japan}
\altaffiltext{8}{Department of Physics and Astronomy, Clemson University, Kinard Lab of Physics, Clemson, SC 29634-0978, USA}
\altaffiltext{9}{Istituto Nazionale di Fisica Nucleare, Sezione di Padova, I-35131 Padova, Italy}
\altaffiltext{10}{Dipartimento di Fisica e Astronomia ``G. Galilei,'' Universit\`a di Padova, I-35131 Padova, Italy}
\altaffiltext{11}{Yale Center for Astronomy and Astrophysics, Physics Department, Yale University, PO Box 208120, New Haven, CT 06520-8120, USA}
\altaffiltext{12}{Cahill Center for Astronomy and Astrophysics, Caltech, Pasadena, CA 91125, USA}
\altaffiltext{13}{Department of Physics, Durham University, Durham DH1 3LE, UK}
\altaffiltext{14}{Universit\'{e} de Toulouse, UPS - OMP, IRAP, Toulouse, France}
\altaffiltext{15}{CNRS, Institut de Recherche en Astrophysique et Plan\'{e}tologie, 9 Av.\ colonel Roche, BP 44346, F-31028 Toulouse Cedex 4, France}
\altaffiltext{16}{Space Science Laboratory, University of California, Berkeley, CA 94720, USA}
\altaffiltext{17}{DTU Space, National Space Institute, Technical University of Denmark, Elektrovej 327, DK - 2800 Lyngby, Denmark}
\altaffiltext{18}{Lawrence Livermore National Laboratory, Livermore, CA 94550, USA}
\altaffiltext{19}{ASI Science Data Center, Via del Politecnico snc I-00133, Roma, Italy}
\altaffiltext{20}{Columbia Astrophysics Laboratory, Columbia University, New York, NY 10027, USA}
\altaffiltext{21}{Core of Research for the Energetic Universe, Graduate School of Science, Hiroshima University, Higashi-Hiroshima, Hiroshima 739-8526, Japan}
\altaffiltext{22}{INAF - Osservatorio Astronomico di Roma, via di Frascati 33, I-00040 Monteporzio, Italy}
\altaffiltext{23}{Jet Propulsion Laboratory, California Institute of Technology, Pasadena, CA 91109, USA}
\altaffiltext{24}{INAF - Osservatorio Astronomico di Brera, Via E, Bianchi 46, I-23807 Merate, Italy}
\altaffiltext{25}{NASA Goddard Space Flight Center, Greenbelt, MD 20771, USA}
\altaffiltext{26}{Harvard-Smithsonian Center for Astrophysics, Cambridge, MA 02138 USA}
\altaffiltext{27}{Hiroshima Astrophysical Science Center, Hiroshima University, Higashi-Hiroshima, Hiroshima 739-8526, Japan}
\altaffiltext{28}{Aryabhatta Research Institute of Observational Sciences (ARIES), Manora Peak, Nainital - 263 002, India}


\begin{abstract}
We report the results of a multi-band observing campaign on the famous blazar 3C~279 conducted 
during a phase of increased activity from 2013 December to 2014 April, including 
first observations of it with \NuSTAR.
The $\gamma$-ray emission of the source measured by \Fermi-LAT showed multiple distinct 
flares reaching the highest flux level measured in this object since the beginning 
of the {\Fermi} mission, with $F(E > 100\,{\rm MeV})$ of $10^{-5}$ photons 
cm$^{-2}$ s$^{-1}$, and with a flux doubling time scale as short as 2 hours.
The $\gamma$-ray spectrum during one of the flares 
was very hard, with an index of $\Gamma_\gamma = 1.7 \pm 0.1$, which is 
rarely seen in flat spectrum radio quasars. 
The lack of concurrent optical variability 
implies a very high Compton dominance parameter $L_\gamma/L_{\rm syn} > 300$.
Two 1-day \NuSTAR\ observations with accompanying \Swift\ pointings 
were separated by 2 weeks, probing different levels of source activity.
While the 0.5$-$70\,keV X-ray spectrum obtained during the first pointing, 
and fitted jointly with \Swift-XRT is well-described by a simple power law, 
the second joint observation showed an unusual spectral structure: 
the spectrum softens by $\Delta\Gamma_{\rm X} \simeq 0.4$ at $\sim4$\,keV.
Modeling the broad-band SED during this flare with the standard synchrotron 
plus inverse Compton model requires: (1) the location of the $\gamma$-ray 
emitting region is comparable with the broad line region radius, 
(2) a very hard electron energy distribution index $p \simeq 1$, 
(3) total jet power significantly exceeding the 
accretion disk luminosity $L_{\rm j}/L_{\rm d} \gtrsim 10$, and (4) extremely
low jet magnetization with $L_{\rm B}/L_{\rm j} \lesssim 10^{-4}$.
We also find that single-zone models that match the observed $\gamma$-ray 
and optical spectra cannot satisfactorily explain the production of X-ray emission. 
\end{abstract}



\section{Introduction}
\label{sec_intro}
 
Blazars are active galaxies where the strong, non-thermal 
electromagnetic emission, generally 
detected in all observable bands from the radio to $\gamma$-ray spectral 
regimes, is dominated by the relativistic jet pointing close to our line of 
sight. Detailed studies of blazar spectra, and in particular the spectral 
variability, are indispensable tools to determine the physical processes responsible 
for the emission from the jet, leading to understanding the distribution 
of radiating particles, and eventually, the processes responsible for their 
acceleration.  

3C~279 is among the best studied blazars;  it is detected in all accessible 
spectral bands, revealing highly variable emission. 
It consistently shows strong $\gamma$-ray emission, 
already clearly detected with the EGRET instrument on the 
{\it Compton Gamma-Ray Observatory} \citep[{\it CGRO};][]{Har92}.  
The object, at $z=0.536$ \citep{Lyn65}, is associated with a 
luminous flat-spectrum radio quasar (FSRQ) with prominent broad 
emission lines.  Optical and UV 
observations in the low-flux state \citep{1999ApJ...521..112P} 
allow the luminosity of the accretion disk to be estimated at 
$L_{\rm d} \simeq 6 \times 10^{45}$ erg s$^{-1}$.\footnote{\cite{1999ApJ...521..112P} report a lower value $L_{\rm d} = 2\times 10^{45}\;{\rm erg\,s^{-1}}$, apparently without a bolometric correction, and we used that value previously in \cite{hay12}. Here, we apply a correction by factor 3.}
The estimates of the 
mass of the central black hole are in the range of $(3 - 8)\times 10^{8}\,M_{\Sun}$, 
derived from the 
luminosity of broad optical emission lines \citep{Woo02}, the 
width of the $H_{\beta}$ line \citep{Gu01}, and 
the luminosity of the host galaxy \citep{2009A&A...505..601N}.  The object possesses 
a compact, milliarcsecond-scale radio core and a jet with time-variable 
structure. Multi-epoch radio observations 
conducted between 1998 and 2001 by \cite{Jor04,Jor05} 
provided an estimate of the bulk Lorentz factor of the radio-emitting 
material, $\Gamma_{\rm j} = 15.5 \pm 2.5$ and the direction of motion 
to the line of sight, $\theta_{\rm obs}=2.1 \pm 1.1$ degrees, which corresponds 
to a Doppler factor $\mathcal{D}$ of $24.1 \pm 6.5$.  

As is the case for blazars, % associated with quasars, 
the most compelling mechanism for the production of the 
radio through optical bands is synchrotron emission, while 
the $\gamma$-rays arise via inverse Compton emission by the same 
relativistic electrons producing the synchrotron emission \citep{2009ApJ...704...38S}. 
Alternative models involving hadronic interactions require significantly higher jet powers due to their lower radiative efficiency \citep{Boe09}.
Since in the co-moving frame of the relativistic jet the photon energy 
density in luminous blazars is dominated by external radiation sources, 
production of $\gamma$-rays is most efficient
by scattering of the external photons \citep{Der92,Sik94}.
3C~279 is regularly monitored by the \Fermi\ satellite together with many different 
facilities covering a range of spectral bands, from radio and optical 
to X-rays. The correlations of the highly variable time series between 
the optical polarization level/angle and $\gamma$-rays provide 
strong evidence for the synchrotron + Compton model, and suggest among the solutions 
that the jet structure is not axisymmetric \citep{Nature}, or the presence of a helical magnetic field component \citep{Zha15}.
The rapid variability, together with the rate of change of the 
polarization angle, suggest a compact (light days) emission region that is 
located at an appreciable ($>$ a parsec) distance along the jet from the 
black hole. Furthermore, the close but not exact correlation of the 
optical and $\gamma$-ray flares, with the optical lagging
the $\gamma$-rays by $\sim 10$ days \citep{hay12},
%(see also Cohen et al. 2014) 
has supported this basic scenario \citep{Jan12}.  

Perhaps the largest mystery in 3C~279 --- and other luminous blazars 
as well --- is the nature of its X-ray emission \citep{2013ApJ...779...68S}.  
Early comparison of the {\it RXTE} X-ray and EGRET $\gamma$-ray time series 
%and McHardy et al. (XX) 
revealed a close association of the $\gamma$-ray 
and X-ray flares~\citep{Weh98}, suggesting that 
the X-ray flux might be the low-energy end of the same inverse Compton emission component
detected at higher energies by EGRET.
This is supported indirectly by a good overall correlation between 
long term {\it RXTE} and optical data (which, according to the above, 
should be a reasonable proxy for $\gamma$-ray flux), although individual 
flares show time lags up to $\sim \pm 20$ days \citep{Cha08}. However, better sampling provided 
by the multi-band time series covering many years \citep[and owing mainly to the 
all-sky monitoring capability of the \Fermi\ Large Area Telescope, LAT;][]{LAT} revealed that 
the $\gamma$-ray and X-ray fluxes are often not well correlated %--- and this might be the case 
both for this object \citep{hay12} and for other blazars \citep[see e.g.,][]{Bon09}. 
The nature of blazar X-ray emission is still somewhat unclear.

3C~279 is also a prominent hard X-ray and soft $\gamma$-ray source, detected by 
{\it CGRO}/OSSE \citep{Har96}, {\it INTEGRAL} \citep{Bec06}, and \Swift-BAT \citep{Tue10}. 
However, these observations did not provide a precise measurement of the hard X-ray 
spectrum of this source, that would allow discrimination between alternative spectral 
components.  3C~279 was selected as one of a few blazar targets to be observed in 
the early phase of the {\it Nuclear Spectroscopic Telescope Array}~\citep[\NuSTAR;][]{NuSTAR} 
focusing hard X-ray (3-79 keV) mission.

After a brief hiatus, 3C~279 became very active late in 2013,
producing a series of $\gamma$-ray flares and 
reaching the highest $\gamma$-ray flux level recorded by the \Fermi-LAT~\citep{Atel} for this source.
The flaring activities of 3C~279 triggered many observations, including the first two 
pointings by the \NuSTAR\ satellite, enabling sensitive spectral measurements 
up to 70\,keV. Here, we present the results of the analysis of the 
\Fermi-LAT, \NuSTAR, and \Swift\ data together with optical 
observations by the SMARTS and Kanata telescopes as well 
as the sub-mm data from the Submillimeter array (SMA) collected for five months 
during the high-activity period, from 2013 November to 2014 April.
A part of this period --- 2014 March$-$April --- was studied independently by \cite{Pal15}.
In Section \ref{sec_data} we describe in detail the data analysis procedures 
and basic observational findings.
In Section \ref{sec_multiband} we compare the observational results 
between multiple bands.
In Section \ref{sec_disc} we discuss the interpretation and theoretical 
implications of our results, and we conclude in Section \ref{sec_conc}.

\begin{figure*}[htbp]
\centering
\includegraphics[height=17.1cm]{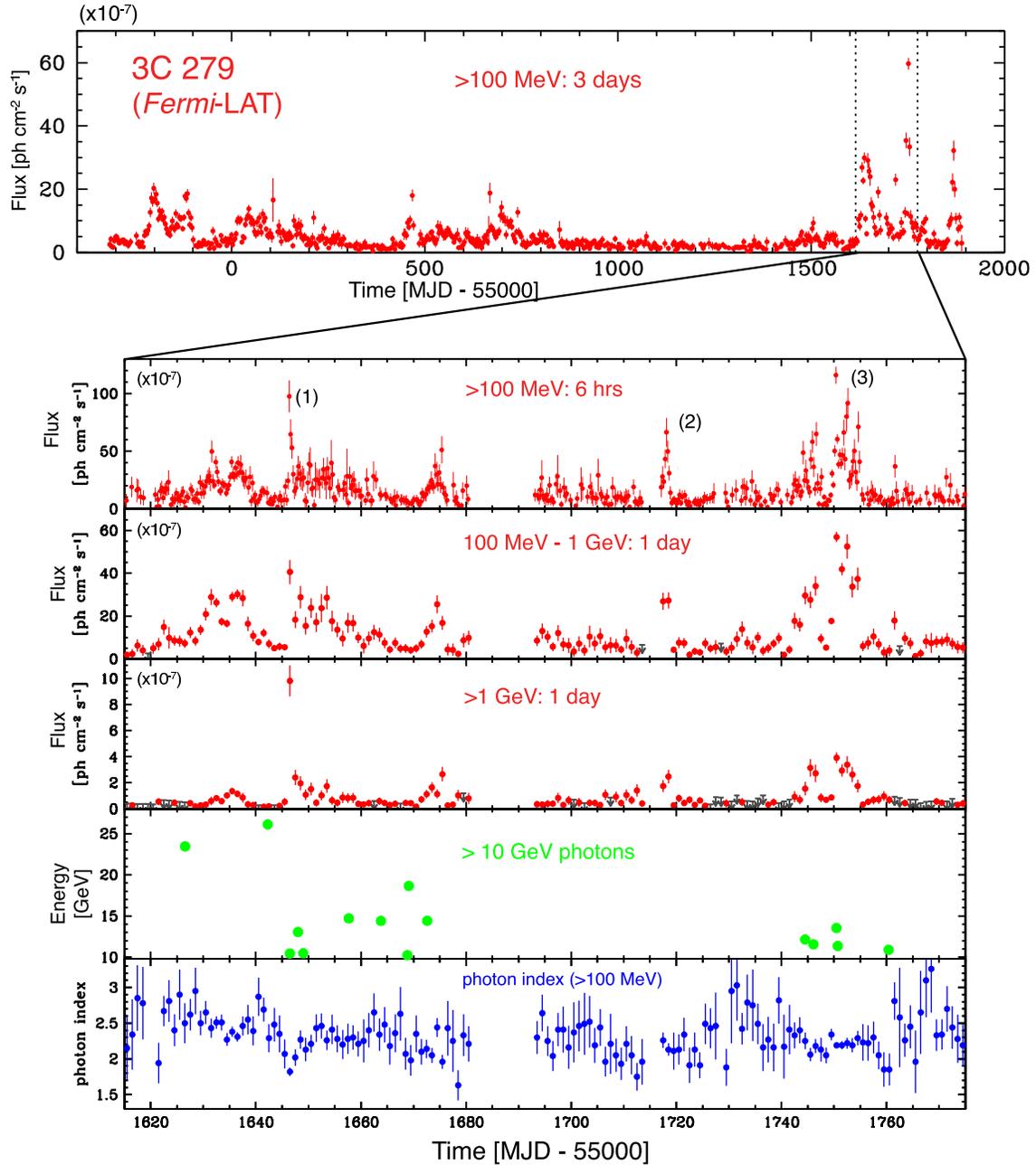}
\caption{Light curves of 3C~279 in the $\gamma$-ray band (integral photon flux) as observed by \Fermi-LAT.
Top panel shows the long-term light curve above 100\,MeV in 3-day bins.
The other panels show light curves for the 2013--2014 active period:
from the top to bottom,
(1) above 100\,MeV in 6-hour bins,
(2) from 100\,MeV to 1\,GeV in 1-day bins, 
(3) above 1\,GeV in 1-day bins,
(4) arrival time distribution of photons with energies above 10\,GeV,
(5) photon index of 3C~279 above 100\,MeV in 1-day bins.
A gap in the data around $\sim$MJD~56680--56690 is due to a ToO observation of the Crab Nebula, during which time no exposure was available in the direction of 3C~279.
The vertical bars in data points represent 1\,$\sigma$ statistical errors and 
the down arrows indicate 95\% confidence level upper limits.
}
\label{LC}
\end{figure*}

\section{Observations and Data Reduction}
\label{sec_data}

\subsection{\Fermi-LAT: Gamma-ray observations}
\label{sec_fermi}

The LAT is a pair-production telescope onboard the \textit{Fermi} satellite with large effective 
area ($\simeq 6500$\,cm$^2$ on axis for 1 GeV photons) and a large field of view (2.4\,sr), 
sensitive from $20$\,MeV to $300$\,GeV 
\citep{LAT}. 
Here, we analyzed LAT data for the sky region including 3C~279  
following the standard 
procedure\footnote{\texttt{http://fermi.gsfc.nasa.gov/ssc/data/analysis/}}, 
using the LAT analysis software \texttt{ScienceTools} \texttt{v9r34v1} with the 
\texttt{P7REP\_SOURCE\_V15} instrument response functions. 
The azimuthal dependence of the effective area was taken into account 
for analysis with short time scales ($<$1 day).
Events in the energy 
range 0.1--300\,GeV were extracted within a 
$15^{\circ}$ acceptance cone of the Region of Interest (ROI) 
on the location of 3C~279 (RA = $195.047^{\circ}$, 
DEC=$-5.789^{\circ}$, J2000). 
It is known that the Sun comes very close to and occults 3C~279 on October 8 each year.
The data when the source is within $5^{\circ}$ of the Sun were excluded.
Gamma-ray fluxes and spectra 
were determined by an unbinned maximum likelihood fit with \texttt{gtlike}. 
We examined the significance of the $\gamma$-ray signal from the sources
by means of the test statistic (TS) based on the likelihood ratio 
test\footnote{$TS=25$ with 2 degrees of freedom corresponds to an estimated $\sim4.6\, \sigma$ pre-trials statistical significance assuming that the null-hypothesis TS distribution follows a $\chi^2$ distribution \citep[see][]{ML}.}.
The background model included all known $\gamma$-ray sources within the ROI 
from the 2-year LAT catalog \citep[2FGL:][]{2FGL}. Additionally, the model included 
the isotropic and Galactic diffuse emission components\footnote{\texttt{iso\_source\_v05.txt} and \texttt{gll\_iem\_v05\_rev1.fit}}.  
Flux normalizations for the diffuse 
and background sources were left free in the fitting procedure.

\begin{figure*}[htbp]
\centering
\plotone{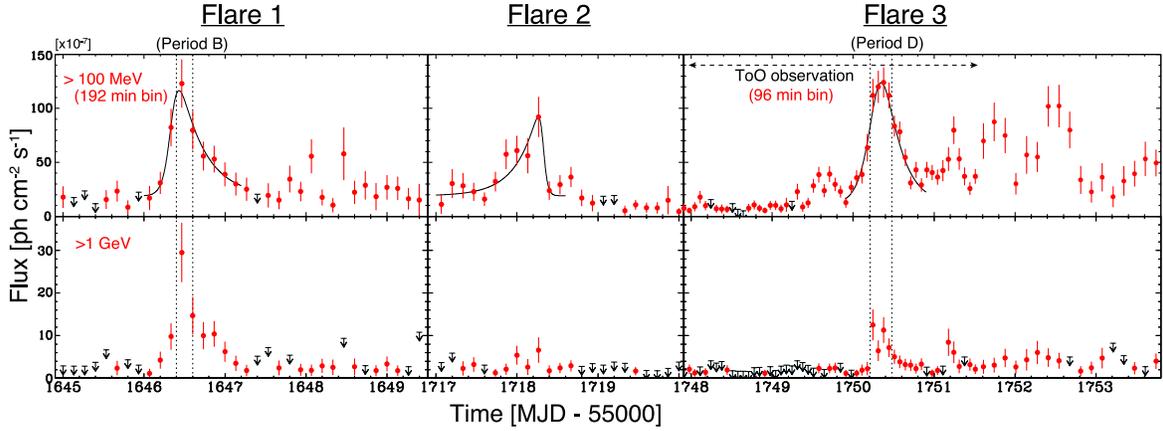}
\caption{Gamma-ray light curves (integral photon flux) of 3C\,279 around the three large flares with fine time bins.
Top panels: $>$ 100\,MeV; lower panel: $>$ 1\,GeV.
For Flares 1 and 2, the bins are equal to two \Fermi\ orbital periods (192 min). 
For Flare 3, during a ToO observation, the bins are equal to one \Fermi\ orbital period (96 min).
The vertical bars in data points represent 1\,$\sigma$ statistical errors and 
the down arrows indicate 95\% confidence level upper limits.
}
\label{shortLC}
\end{figure*}

\begin{deluxetable*}{c|ccccc}
\tablecaption{Fitting results of the light curve profile in the $\gamma$-ray band measured by \Fermi-LAT.\label{tab:LCfit}}
\tabletypesize{\small}
\centering                         
\startdata       
\hline \hline 
Flare & $\tau_{\rm rise}$ & $\tau_{\rm fall}$ & $b$ &$F_0$  & $t_0$ \\
number & (hrs) & (hrs) & ($10^{-7}$\,\phcms) & ($10^{-7}$\,\phcms)  & (MJD) \\
\hline 
Flare 1 & $1.4\pm0.8$ & $7.4\pm3.2$ & $ 150\pm36$  &  $19\pm12$  & $56646.35\pm0.04$\\
Flare 2 & $6.4\pm2.4$  & $0.68\pm0.59$ & $100\pm26$ &  $19\pm5$ & $56718.32\pm0.07$ \\
Flare 3 (ToO) & $  2.6\pm0.6$ & $5.0\pm0.8$ & $216\pm19$ & $10.5\pm6.6$  & $56750.30\pm0.04$  
\enddata
\end{deluxetable*}

Several $\gamma$-ray light curves as measured by \Fermi-LAT can be seen in Figure~\ref{LC}.
The top stand-alone panel shows the $\gamma$-ray flux above 100\,MeV for about 
6 years since the beginning of scientific operations of the \Fermi-LAT (2008 August 5) 
up to 2014 August 31 (MJD~54683--56900) binned into 3-day intervals.
After $\sim$MJD~56600, the source entered the most active state since the launch of \Fermi\ satellite.
This resulted in 
Target-of-Opportunity (ToO) pointing observations for 3C~279, which 
were performed 
between 2014 March 31 21:59:47 UTC (MJD~56747.91652) and 2014 April 04 12:42:01 UTC (MJD~56751.52918), and those observations are included in our analysis.  
The time series of the $\gamma$-ray flux and photon index of 3C~279 measured
with \Fermi-LAT during the most active states from MJD 56615 (2013 November 19) to MJD 56775 (2014 April 28),
are illustrated in other panels in Figure~\ref{LC}.

Three distinct flaring intervals are evident in the $\gamma$-ray light curve:
Flare 1 ($\sim$MJD\,56650), Flare 2 ($\sim$MJD\,56720) and Flare 3 ($\sim$MJD\,56750). 
The maximum 1-day averaged flux above 100\,MeV reached 
$(62.2\pm2.4)\times10^{-7}$\,{\phcms} (${\rm TS}=3892$) on MJD\,56749 (2014 April 03)
\footnote{throughout this paper, each error represents a $1\,\sigma$ statistical uncertainty},
which is about three times higher than the maximum 1-day
 averaged flux recorded during the first two years~\citep[on MJD 54800:][]{hay12}.
 On the other hand, the maximum 1-day averaged flux above 1\,GeV was observed 
on MJD\,56645 (2013 December 20) at $(9.8\pm1.2)\times10^{-7}$\,\phcms,
much higher than the $>1$\,GeV flux on MJD\,56749, which was $(3.9\pm0.4)\times10^{-7}$\,\phcms.
The photon index also shows a hardening trend toward MJD\,56645, when it reached a very hard index of $1.82\pm0.06$, 
which is rarely observed in FSRQs.

Figure~\ref{shortLC} shows detailed light curves around the flares with short time bins.
During Flares 1 and 2, the fluxes were derived with an interval of 192 min, 
corresponding to two orbital periods of \Fermi-LAT. 
During Flare 3, because the ToO pointing to 3C~279 increased the exposure, time bins as short as one orbital period (96 min) were used. 
The peak flux above 100\,MeV in those time intervals (192\,min and 96\,min) reached $\sim 120\times 10^{-7}$\,\phcms\ .

The very rapid variability apparent in the data can be fitted 
by the following function to characterize the time profiles of the source flux variations:  
\begin{equation}
F(t) = F_0 + \frac{b}{e^{-(t-t_0)/\tau_{\rm rise}}+ e^{(t-t_0)/\tau_{\rm fall}}}
\label{eq:timep}
\end{equation}
This formula has also been used in variability studies of other LAT-detected bright blazars 
to characterize the temporal structure of $\gamma$-ray light curves~\citep{vari}.
The double exponential form 
has been applied previously to the light curves of
blazars~\citep{Val99} as well as 
gamma-ray bursts~\citep[e.g.,][]{Nor00}.
In this function, each $\tau_{\rm rise}$ and $\tau_{\rm fall}$ represents the 
"characteristic" time scale for the rising and falling parts of the light curve, respectively,
and $t_0$ describes approximatively the time of the peak (it corresponds to the actual
maximum only for symmetric flares). 
In general, 
the time of the maximum of a flare ($t_p$) can be described using parameters in Equation~\ref{eq:timep} as:
\begin{equation}
t_p = t_0 + \frac{\trise \tfall}{\trise+\tfall}\ln{\left( \frac{\tfall}{\trise} \right)}
\end{equation}
The parameters of the fitting results are summarized in Table~\ref{tab:LCfit}.
The time profiles show asymmetric structures in all flares; generally 
the rise times correspond to 1-2 hours, which are several times shorter than 
the fall times of 5-8 hours in Flares 1 and 3. On the other hand, the fall 
time appears to be less than 1 hour in Flare~2 (although the fitting error of the 
parameter is quite large). One can see in the light curve of Flare 2 in 
Figure~\ref{shortLC} that the flux reached $\sim 90\times10^{-7}$\,\phcms 
at the peak but suddenly dropped by a factor of $\sim$~3 in the next bin, 2 orbits (196 min) later.

\begin{deluxetable*}{c|cccccccc}
\tablecaption{Results of spectral fitting in the $\gamma$-ray band measured by \Fermi-LAT.\label{tab:GammaSPfit}}.
\tabletypesize{\scriptsize}
\centering                         
\startdata       
\hline \hline
Period & \multicolumn{6}{c}{Gamma-ray spectrum (\Fermi--LAT) } & Flux ($>0.1$\,GeV) & \# of photons \\
(MJD - 56000) & fitting model\tablenotemark{a} & $\Gamma_\gamma/\alpha/\Gamma_{\gamma 1}$ & $\beta/\Gamma_{\gamma 2}$ &   $E_{\rm brk}$ (GeV) &  $TS$ & $-2\Delta{L}$\tablenotemark{b} & $ (10^{-7}$ \phcms) & $>10$\,GeV  \\
\hline 
Period A (3 days)  &  PL & $2.36\pm0.13$ & \nodata & \nodata & 174  & \nodata & $5.9\pm0.9$ & 1 \\
Dec.\,16,\,0h -- 19,\,0h &  LogP  & $2.32\pm0.17$ & $0.03\pm0.07$ & \nodata & 174  & $<0.1$  & $5.7\pm0.9$ & (26.1\,GeV)\\
($642.0-645.0$) &  \\
\hline 
Period B (0.2 days)  &  PL & $1.71\pm0.10$ & \nodata & \nodata & 407  & \nodata & $117.6\pm19.7$ & 1\\
Dec.\, 20,\,9h36 -- 14h24 &  LogP  & $1.12\pm0.31$ & $0.19\pm0.09$ & \nodata & 413  & 6.0 & $94.5\pm18.1$ & (10.4\,GeV)\\
 ($646.4-646.6$)   &  BPL  & $1.41\pm0.17$ & $3.01\pm0.91$ & $3.6\pm1.6$ & 415  & 7.6 & $100.6\pm18.4$ \\
\hline 
Period C (3 days)  &  PL & $2.29\pm0.13$ & \nodata & \nodata & 219  & \nodata & $17.1\pm2.8$  & 1 \\
Dec.\,31,\,0h -- Jan.\,03,\,0h  &  LogP  & $2.29\pm0.16$ & $0.00\pm0.06$ & \nodata & 219  & $<0.1$  & $17.1\pm2.9$ & (14.7\,GeV) \\
  ($657.0-660.0$) &  BPL  & $2.22\pm0.42$ & $2.32\pm0.20$ & $0.34\pm0.27$ & 219  & $<0.1$ & $16.9\pm3.1$ & \\
\hline 
Period D (0.267 days)  &  PL & $2.16\pm0.06$ & \nodata & \nodata & 1839  & \nodata & $117.9\pm7.1$  & 1 \\
Apr.\,03,\,5h03 -- 11h27  &  LogP  & $2.02\pm0.08$ & $0.10\pm0.05$ & \nodata & 1840  & $ 5.3$  & $114.9\pm7.1$ & (13.5\,GeV)\\
  ($750.210-750.477$) &  BPL  & $2.02\pm0.09$ & $2.89\pm0.45$ & $1.6\pm0.6$ & 1843  & $8.0$ & $115.1\pm7.7$ 
\enddata
\tablenotetext{a}{PL: power law model, LogP: log parabola model, BPL: broken power law model. See definitions in the text.}
\tablenotetext{b}{ $\Delta L$ represents the difference of the logarithm of the total likelihood of the fit with respect to the case with a PL for the source.}
\end{deluxetable*}

\begin{figure}[tbp]
\centering
\includegraphics[width=\columnwidth]{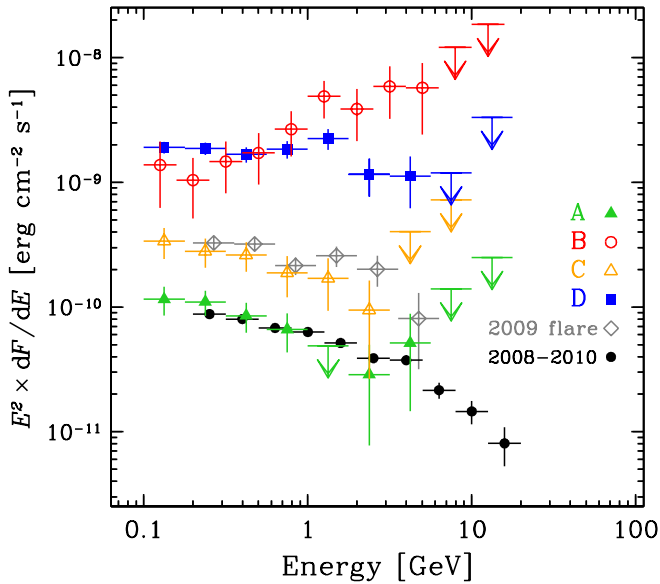}
\caption{Gamma-ray spectral energy distribution of 3C\,279 as measured by \Fermi-LAT during the four periods identified in the text (see Section~\ref{sec_fermi}) as well as in Table~\ref{tab:GammaSPfit}. 
The plot includes the spectra of 3C\,279 from the 2008--2010 campaign \citep{hay12}, including a large flare and a two-year average.
In data points, the horizontal bars describe the energy ranges of bins and
the vertical bars represent 1\,$\sigma$ statistical errors. 
The down arrows indicate 95\% confidence level upper limits.  
}
\label{GammaSP}
\end{figure}

Gamma-ray spectra were extracted from the following four periods: 

\begin{itemize} 

\item (A) Overlapping with the first \NuSTAR\ observation (see Section~\ref{sec:NuSTAR}). 
Although the \NuSTAR\ observation lasted for about one day, 
in order to increase the $\gamma$-ray photon statistics, 
the LAT spectrum was extracted from 3 days where the source showed 
comparable flux level (as inferred from the light curve with 1-day bins).  
In this period, the source was found to be in a relatively low state.

\item (B) For three orbits ($\sim 4.5$ hours) at the peak of Flare 1, 
when the source showed a very hard $\gamma$-ray photon index ($<$2).

\item (C) Overlapping with the second \NuSTAR\ observation (see Section~\ref{sec:NuSTAR}).
As in the case of Period A, the length of this period is 3 days,
while the \NuSTAR\ observation lasted about 1 day.
The source flux was higher than in Period A.

\item (D) At the peak of Flare 3 for 4 orbits ($\sim 6$ hours).

\end{itemize}

In a similar manner to previous spectral studies of 
the source with the \Fermi-LAT~\citep{hay12, ale14}, 
each $\gamma$-ray spectrum was modeled using a simple 
power-law (PL; ${\rm d}N/{\rm d}E \propto E^{-\Gamma_\gamma}$), 
a broken power-law (BPL; ${\rm d}N/{\rm d}E \propto E^{-\Gamma_{\gamma 1}}$ 
for $E<E_{\rm brk}$ 
and ${\rm d}N/{\rm d}E \propto E^{-\Gamma_{\gamma 2}}$ otherwise), 
and a log-parabola model (LogP; ${\rm d}N/{\rm d}E \propto (E/E_0)^{-\alpha - \beta \log(E/E_0)}$, with $E_0=300$\,MeV).
The spectral fitting results are summarized in Table~\ref{tab:GammaSPfit}
and a spectral energy distribution for each period is plotted in Figure~\ref{GammaSP}.
In contrast to the general feature of FSRQs that the photon index is almost 
constant regardless of the source flux~\cite[see e.g.,][]{hay12},
the spectral shape significantly changed between the periods.
Remarkably, the photon index of the simple PL model for the Period B 
resulted in an unusually hard index for FSRQs, of $1.71\pm0.10$.
Such a hard photon index has not been previously reported 
in past LAT observations of 3C~279 that included several flaring 
episodes~\citep{hay12, ale14}.
Among the sources in the Second LAT AGN Catalog~\citep[2LAC:][]{2LAC}, 
the mean photon index value of FSRQs is 2.4, and 
only one FSRQ (2FGL~J0808.2$-$0750) in the Clean and flux-limited Sample 
has the photon index of $\Gamma_\gamma < 2$~\citep[see Figure~18 in][]{2LAC}.
Occasionally, hard photon indices have been observed in bright FSRQs during rapid flaring events~\citep{Pac14}. 
The photon index of Period B is even harder than the index
of 4C\,+21.35~\citep[$1.95\pm0.21$;][]{ale11} and
of PKS\,1510$-$089~\citep[$2.29\pm0.02$;][]{1510MAGIC}
at the time when the $>100$\,GeV emission was detected.

No significant deviations from a PL model were detected in the spectra of Period A and C,
while evidence of spectral curvature was observed in the spectra of the flare peaks, Period B and D.
As derived fitting the BPL model for Period B, the photon 
index of the lower energy part (below $E_{\rm brk} = 3.6 \pm 1.6$ GeV) is $1.41 \pm 0.17$.
This is comparable to the photon index of the rising part of the inverse-Compton 
emission for the case of a parent electron index of 2, as is typical of the $\gamma$-ray 
spectra of high-frequency peaked BL Lac objects.
One can easily recognize such a rising spectral feature in Figure~\ref{GammaSP}.
On the other hand, Period D also shows a very high flux, exceeding $10^{-5}$\,\phcms ($>$ 100\,MeV), 
comparable to the flux of Period B.
However, spectral shape is characterized by a soft index ($\Gamma_\gamma > 2$).
The photon index of the lower energy part as derived from fitting with the BPL model is not significantly harder than 2,
nor is a rising spectral feature apparent in the SED plot of Figure~\ref{GammaSP}.

\subsection {X-ray observations}
\subsubsection{\NuSTAR: hard X-rays}
\label{sec:NuSTAR}

\NuSTAR\ is a Small Explorer satellite sensitive to hard X-rays, covering 
the bandpass of 3 -- 79\,keV. It features two multilayer-coated optics, 
focusing the reflected X-rays onto CdZnTe pixel detectors which 
provide spectral resolution (FWHM) of 0.4\,keV at 10\,keV, increasing to 0.9\,keV at 68\,keV.
The field of view of each 
telescope is $\sim 13'$, and the half-power diameter of the point spread 
function is $\sim 1'$.  The low background resulting from 
focusing of X-rays provides an unprecedented sensitivity for measuring fluxes and spectra of 
celestial sources. For more details, see \citet{NuSTAR}.

\NuSTAR\ observed 3C~279 twice. 
The first observation was performed between 2013 December 16, 05:51:07 and 2013 December 17, 04:06:07 (UTC)
and the second one between 2013 December 31, 23:46:07 and 2014 January 01, 22:11:07 (UTC).
The raw data products were processed with the 
\NuSTAR\ Data Analysis Software (NuSTARDAS) package v.1.3.1,
using the {\tt nupipeline} software module which produces calibrated and cleaned event files.
We used the calibration files available in the NuSTAR CALDB calibration data base v.20140414.
Source and background data were 
extracted from a region of $1\farcm5$ radius, centered respectively 
on the centroid of the X-ray source, and a region $5'$ N of the source 
location on the same chip. Spectra were binned in order to over-sample the instrumental 
resolution by at least a factor of 2.5 and to have
a signal-to-noise ratio greater than 4 in each spectral channel. 
Net ``on source'' exposure times
corresponded to $\sim 39.6$\,ks and $\sim 42.7$\,ks for 
the first (Dec.~16) and second (Dec.~31) observations, respectively.
We considered the spectral channels corresponding nominally to the 3.0 -- 70 
keV energy range. The net (background subtracted) count rates for the first 
observation were $0.303 \pm 0.003$ and $0.294 \pm 0.003$ cnt s$^{-1}$ 
respectively for module A and module B, while for the second observation 
they were $0.636 \pm 0.004$ and $0.590 \pm 0.004$ cnt s$^{-1}$. 
We plotted the raw (not background-subtracted) counts 
binned on an orbital time scale in Figure~\ref{XrayLC}. 
It is apparent that the source was variable from one observation to the other, but also that the 
source varied within the second observation via secular decrease of 
flux for the first 3 hours, followed by an increase by nearly 
a factor of 2.

The spectral fitting was performed using XSPEC v12.8.1 with the standard instrumental response 
matrices and effective area files derived 
using the NuSTARDAS software module {\tt nuproducts}.
For each observation, we fitted the two modules simultaneously 
including a small normalization factor for module B with respect to 
the module A in the model parameters.
We adopted simple power-law and a broken-power-law models modified
by the effects of the Galactic absorption, corresponding to a column of 
$2.2 \times 10^{20}$ cm$^{-2}$~\citep{Kal05}.
The results of the two spectral fits were compared against each other by using an F-test  
to examine improvements by the broken-power-law model.
The simple power-law model gave acceptable results for both observed spectra, 
with $\chi^2/{\rm d.o.f.}$ of 666.8/660 (41.9\% for the corresponding $\chi^2$ probability) and 831.2/886 (90.6\%), respectively.
Although the model fluxes in the 2--10 keV band between the two observations showed a difference of about a factor of 2,
the resulting photon indices were similar: $1.739\pm0.013$, and $1.754\pm0.008$. 
While there was no improvement in the fit obtained using the broken-power-law model in the first observation,
it gave slightly 
better fits for the spectrum of the second observation, 
with a $\chi^2$/d.o.f.\ of 820.8/884 (93.6\%), 
yielding a probability of 0.4\% ($\sim 2.9\,\sigma$) that the improvement in the fit was due to chance 
(as assessed with an F-test).
This may indicate a deviation from a single power law in the spectrum of the second observation.
The fitting results for the \NuSTAR\ spectra are summarized in Table~\ref{tab:Xfit}.

\subsubsection{\Swift-XRT: X-ray}

The publicly available \Swift\ X-Ray Telescope (XRT) data in the HEASARC 
database\footnote{\url{http://heasarc.gsfc.nasa.gov/cgi-bin/W3Browse/swift.pl}} 
reveal that \Swift\ observed 3C~279 51 times between 2013 November and 2014 April.  
We analyzed all those observation IDs (ObsIDs).  
The exposure times ranged from 265\,s (ObsID:35019120) to 9470\,s (ObsID:35019100).
The XRT was used in photon counting mode, and no evidence of pile-up was found. 
The XRT data were first calibrated and cleaned with standard 
filtering criteria with the {\tt xrtpipeline} software module distributed 
with the XRT Data Analysis Software (XRTDAS, version 2.9.2). 
The calibration files available in the version 20140709 of 
the \Swift-XRT CALDB were used in the data reduction.
The source events were 
extracted from a circular region, 20 pixels (1 pixel $\simeq$ 2\farcs36) in radius, centered on the source position.
The background was determined using data 
extracted from a circular region, 40 pixels in radius, centered on 
(RA, Dec: J2000) = (12h56m26s,-05\degree49\arcmin30\arcsec), where no X-ray sources are found.
Note that the background contamination is less 
than 1\,\% of source flux even in the faint X-ray states of the source.
The data were rebinned to have at least 25 counts per bin, and the spectral 
fitting was performed using the energy range above 0.5\,keV using {\tt XSPEC}~v.12.8.1.
The Galactic column density was fixed at $2.2\times10^{20}$ cm$^{-2}$. 
The data were analyzed and the flux and photon index were derived separately for each ObsID.
A relation between the unabsorbed model flux (0.5--5\,keV) and photon index is 
represented in Figure~\ref{SwiftXray}. Only the ObsIDs with an exposure 
of more than 600\,s and with more than 7 spectral points (= d.o.f.~$\geq$~5) were selected for the plot.
A trend of a harder spectrum when the source is brighter is clearly detected.

\begin{figure*}[t]
\centering
\plotone{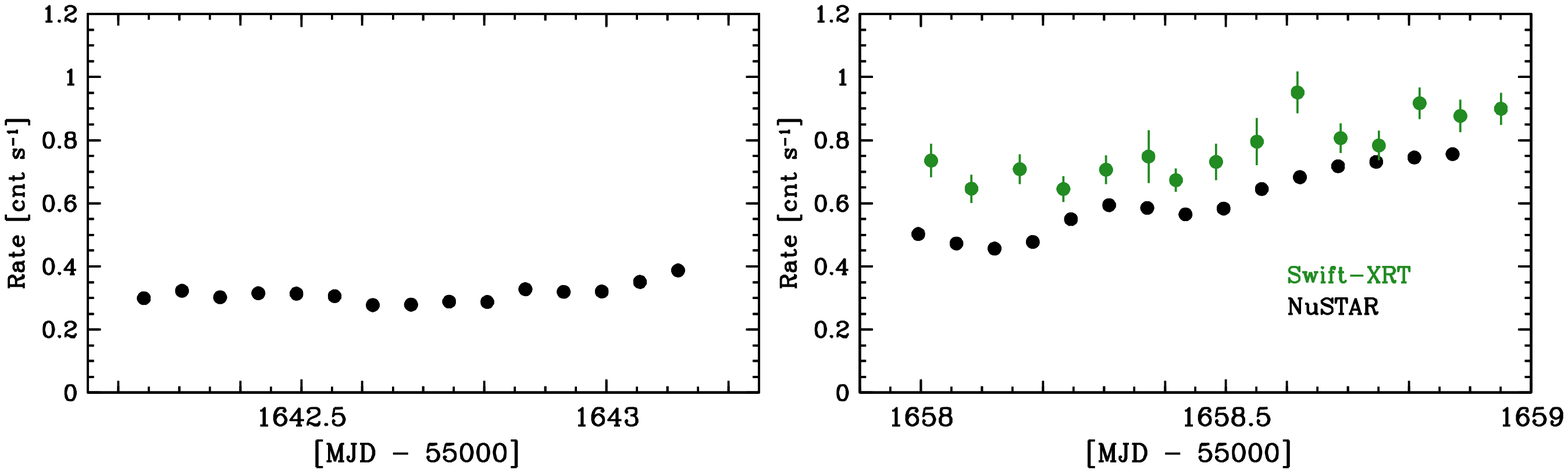}
\caption{X-ray light curves based on the count rates as measured by \NuSTAR\ (black) and by \Swift-XRT (green).
\NuSTAR\ data are plotted in 1.5\,hr bins,
and the \Swift-XRT data are plotted for each snapshot.
The vertical bars represent 1\,$\sigma$ statistical errors.
}
\label{XrayLC}
\end{figure*}

\begin{figure}[htbp]
\centering
\includegraphics[width=0.91\columnwidth]{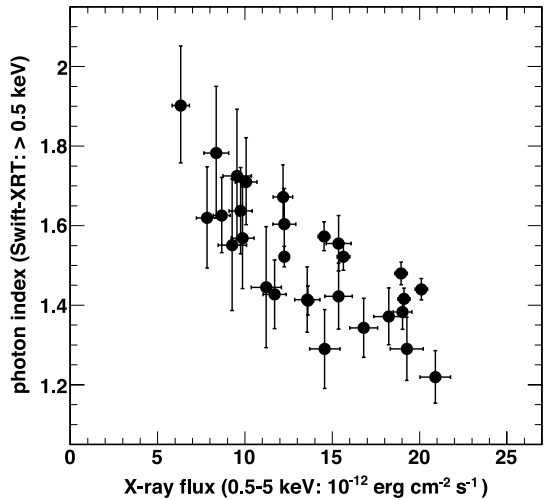}
\caption{Scatter plot of the X-ray flux (0.5--5\,keV) vs.\ the X-ray photon index $\Gamma_{\rm X}$ of 3C~279 based on the \Swift-XRT data. The horizontal and vertical bars describe 1\,$\sigma$ statistical errors for each axis.
}
\label{SwiftXray}
\end{figure}

Data with ObsID of 80090001 and 35019132 were taken (quasi-) 
simultaneously with the \NuSTAR\ observations, and here we report the details of those \Swift-XRT observations.
The observation with ObsID:80090001 was performed from 2013 December 17 21:06:51 to 22:39:56 (UTC) 
with an exposure time of 2125 sec.
Therefore, this observation did not exactly overlap with the \NuSTAR\ observation, 
but it was the closest available, starting about 17 hours after the end of the first \NuSTAR\ observation of the source.
The spectral fit with a power-law model yielded a photon index of $1.67\pm0.08$ 
with a flux in the 0.5--5 \,keV band of $(12.2 \pm 0.6)\times 10^{-12}$\,\erg.

The other observation, with ObsID:35019132, was performed 
between 2014 January 01 00:20:26 and 22:50:54 (UTC), 
which overlaps well 
with the second \NuSTAR\ observation.
The exposure time of this \Swift\ observation was 6131\,s and 
the best-fit power-law model displayed a photon index of $1.42\pm0.03$ 
with a flux in the 0.5--5\,keV band of $(19.1 \pm 0.3)\times 10^{-12}$\,\erg.
The spectral fitting results are reported in Table~\ref{tab:Xfit}.
Each snapshot observation during this ObsID was also analyzed separately.
There were 15 snapshots in total and the exposure time in each snapshot was about 400\,s typically, ranging from about 312\,s to 594\,s. The resultant count rates for all channels are shown in Figure~\ref{XrayLC}.

\subsubsection{Joint spectral fit of \NuSTAR\ and Swift-XRT data}

Joint spectral fits of the \NuSTAR\ and \Swift-XRT data were performed for each \NuSTAR\ observation.
As described in previous sections, the data used for the spectral fitting 
were above 0.5 keV for the \Swift-XRT and 3--70 keV for \NuSTAR.
Here, we introduced a normalization factor of order $(1-3)\%$ with respect to \NuSTAR\ module A
to account for differences in the absolute flux calibrations, and also to account 
for the offsets of the \NuSTAR\ and \Swift\ observing times (necessary 
for an analysis of a variable source). The Galactic column density was fixed 
at $2.2\times10^{20}$ cm$^{-2}$ as above. Simple power-law, a broken power-law 
and a double-broken power-law (for the second observation) models were used for the source 
spectral models. The joint spectral fitting results are summarized in Table~\ref{tab:Xfit}.

\begin{figure}[htbp]
\centering
\includegraphics[width=\columnwidth]{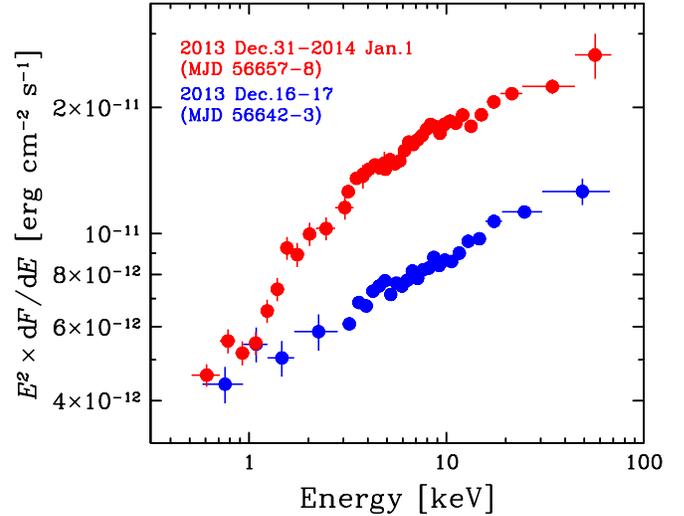}
\caption{Spectral energy distributions of 3C~279 in the soft-hard X-ray band based on the combined data from \Swift-XRT and \NuSTAR.
The blue points show the results for observations on 2013 December 16--17 (in Period A),
and the red points show the results for observations on 2013 December 31 -- 2014 January 01 (in Period C).
The horizontal bars in data points describe the energy ranges of bins 
while the vertical bars represent 1\,$\sigma$ statistical errors.
}
\label{JointXraySED}
\end{figure}

The joint spectrum during the first \NuSTAR\ observation (Dec.~16) can be 
represented by a simple power law from 0.5 to 70 keV with a photon index 
of $1.74\pm0.01$ ($\chi^2/{\rm d.o.f.} = 688.6/679$). The normalization 
factor of \Swift-XRT with respect to \NuSTAR\ module A is $1.01\pm0.05$.
The broken-power-law model 
improved the fit only marginally ($\sim1\,\sigma$) with respect to the power-law model.
The result indicates that the X-ray 
spectrum of 3C\,279 can be described by a single power law from the soft (0.5\,keV) 
to the hard X-ray (70\,keV) band,
which is supported by the results from the individual fits for each \Swift-XRT and \NuSTAR\ observation alone.

\begin{deluxetable*}{ccccccrccc}
\tablewidth{18.3cm}
\tablecaption{Parameters of the spectral fits in X-ray band. \label{tab:Xfit}}
\tabletypesize{\scriptsize}
\centering                         
\startdata     
\hline \hline 
Instrument & $\Gamma_{\rm X}/\Gamma_{\rm X1} $ &  $E_{\rm brk1}$ & $\Gamma_{\rm X2}$ &  $E_{\rm brk2}$ & $\Gamma_{\rm X3}$   & \multicolumn{1}{c}{const.} & $F_{2-10{\rm keV}}$ &  $\chi ^2$/d.o.f. & F-test\\  
(1)  & (2) & [keV] (3)  & (4)  &  [keV] (5) & (6) & \multicolumn{1}{c}{XRT/module B (7)} &  (8) & (9) & (prob.)\\
\hline 
\multicolumn{7}{l}{Data on 2013 December 16 -- 17 (in Period A)} \\
\hline
 \Swift-XRT only &   $1.67\pm0.08$ & \nodata & \nodata &  \nodata &  \nodata &  \nodata/\nodata & $11.9$ & 21.10/18 (27.4\%) \\
\NuSTAR\ only & $1.74\pm0.01$  & \nodata & \nodata &  \nodata &  \nodata &  \nodata/$1.06\pm0.01$ & $11.0$  &  666.8/660 (41.9\%) \\
XRT $+$ \NuSTAR\ & $1.74\pm0.01$ & \nodata & \nodata & \nodata & \nodata &  $1.01\pm0.05/1.06\pm0.01$ & 11.0 & 688.6/679 (39.1\%) & -- \\
 & $ 1.65_{-0.08}^{+0.06}$ & $4.5\pm0.7$ & $1.75\pm0.02$ & \nodata & \nodata & $ 1.11_{-0.08}^{+0.09}/ 1.06\pm0.01$  & 12.0 &  686.2/677  (39.5\%) & 30\% $(\sim1.0\,\sigma)^{\rm a}$ \\
\hline
\multicolumn{7}{l}{Data on 2013 December 31 -- 2014 January 1 (in Period C)} \\
\hline
 \Swift-XRT only &   $1.42\pm0.03$ & \nodata & \nodata &  \nodata  &    \nodata  &  \nodata/\nodata & $23.7$ &  113.1/109 (37.5\%) \\
  \NuSTAR\ only &  $1.75\pm0.01$ & \nodata & \nodata &  \nodata  &  \nodata & \nodata/$1.00\pm0.01$ & $23.4$ &  831.2/886 (90.6\%) & -- \\
 &  $1.71\pm0.02$  & $9.3_{-1.3}^{+1.6}$ & $1.81\pm0.02$ &  \nodata  &  \nodata & \nodata/$1.00\pm0.01$ & $23.2$ &  820.8/884 (93.6\%) & 0.4\% $(\sim2.9\,\sigma)^{\rm a}$ \\
XRT $+$ \NuSTAR\ & $1.73\pm0.01$ &  \nodata & \nodata & \nodata &\nodata & $ 0.70\pm0.02/1.00\pm0.01 $   & \nodata& 1072.4/996 (4.6\%) \\  
&  $1.37\pm0.03 $ &  $3.7\pm0.2$ & $1.76\pm0.01$ & \nodata & \nodata &  $0.97\pm0.03/1.00\pm0.01$ & 22.6 &  940.6/994 (88.6\%) & --\\
 &  $1.37^{+0.03}_{-0.04}$ & $3.6^{+0.2}_{-0.4}$ & $1.72\pm0.02$ & $9.4^{+2.1}_{-1.4}$ & $1.81\pm0.02$ & $0.97\pm0.03$/$1.00\pm0.01$ & 22.6  &  933.4/992 (90.8\%) & 0.22\% $(\sim3.1\,\sigma)^{\rm b}$ \\
 (log parabola)$^{\rm c}$ & $1.39\pm0.03^{\rm d}$ & 1 (fixed)$^{\rm e}$ & $0.19\pm0.02^{\rm f}$ & & & $0.90\pm0.03$/$1.00\pm0.01$ & 22.7 & 960.6/995 (77.8\%) 
\enddata
\tablecomments{Col. (1): Instrument providing the data. Col. (2): Photon index for the power law model, or low-energy photon index for the broken power law model. Col. (3): Break energy [keV] for the broken power law model. Col. (4): High-energy photon index for the broken power law model. Col. (5): Second break energy.  Col (6): Third index in the double-broken power law model.  Col. (7): Constant factor of \Swift-XRT/\NuSTAR\ module-B data with respect to the \NuSTAR\ module-A data. Col. (8): Unabsorbed model flux in the 2-10 keV band, in units of $10^{-12}$ [erg cm$^{-2}$\ s$^{-1}$]. Col. (9): $\chi^2/$degrees-of-freedom and a corresponding probability. \\ 
a: compared to the simple-power-law model. b: compared to the broken-power-law model. c: {\tt logpar} model in XSPEC. d: slope at the pivot energy. e: fixed pivot energy. f: curvature term.
}
\end{deluxetable*}

For the joint spectrum during the second \NuSTAR\ observation, the simple power-law 
model did not result in an acceptable fit, with $\chi^2/{\rm d.o.f.}= 1072/996$.  
Moreover, the normalization factor of \Swift-XRT against the \NuSTAR\ module A 
was $\sim0.7$, which is clearly unacceptable. The broken power-law model, 
on the other hand, gave acceptable results with $\chi^2/{\rm d.o.f.} = 940.6/994$,
and the normalization of the \Swift-XRT data was $0.97\pm0.03$.  
The break energy corresponded to $3.7\pm0.2$\,keV with photon indices 
of $1.37\pm0.03$ and $1.76\pm0.01$, respectively, below and above the break energy.

This break energy is located where the bandpasses of \Swift-XRT 
and \NuSTAR\ data overlap, corresponding respectively 
to the higher and the lower energy end of those data sets.  We are confident that 
the spectral break is a real feature for the following reasons. 
There is a significant difference in the photon 
index in each \Swift-XRT and \NuSTAR\ data set considered individually. This supports the conclusion that 
there is a spectral break at an energy close to the overlap of 
the \Swift-XRT and \NuSTAR\ bandpasses. Each resultant individual photon index is similar 
to the photon index derived from the joint fit below and above the 
break energy, respectively.
The exposures of \Swift-XRT and \NuSTAR\ 
significantly overlapped (see Figure~\ref{XrayLC}), 
yielding a reasonable inter-calibration constant ($0.97\pm0.03$).

We also investigated a double-broken power-law model. 
The model yielded a probability of 
0.22\% ($\sim3.1\sigma$) that the improvement 
in the fit was due to chance against the broken power-law model as assessed with an F-test.
The second break energy appeared at $9.4^{+2.1}_{-1.4}$\,keV and the photon index became 
even softer above the second break energy, changing from $1.72\pm0.02$ to $1.81\pm0.02$.
The spectral break at that energy was also seen in the fitting result 
for the \NuSTAR\ data only at $9.3^{+1.6}_{-1.3}$\,keV.  All these results 
suggest that the X-ray spectrum during the high state (the second \NuSTAR\ 
observation) gradually softens with increasing energy, with the 
photon index changing by $\sim 0.4$ from $\sim 0.5$\,keV to $\sim 70$\,keV.  
This is the first time that detailed, broad-band spectral X-ray measurements 
of 3C~279 show spectral softening with increasing energy. Furthermore 
the {\sl absence} of spectral softening in the first (Dec.~16) \NuSTAR\ observation 
clearly rules out the spectral shape as being caused by additional absorption.  
The joint X-ray spectral data points from the soft to the hard X-ray 
bands obtained by \Swift-XRT and \NuSTAR\ are plotted in 
Figure~\ref{JointXraySED} in the $E \times F(E)$ (erg$\,{\rm cm}^{-2}\,{\rm s}^{-1}$) form.
Finally, a log parabola model was also tested using the {\tt logpar} model in XSPEC.
The pivot energy was fixed at 1\,keV and the best-fit parameters are summarized in Table~\ref{tab:Xfit}.
The model also gave us acceptable fitting results, with $\chi^2/{\rm d.o.f.}= 960.6/995$.

\subsection{UV-Optical observations}

\subsubsection{Swift-UVOT: UV bands}

The \Swift-Ultra-Violet/Optical Telescope (UVOT) data used in
this paper included all observations performed during 
the time interval from 2013 November to 2014 April.
The UVOT telescope cycled through each of the six optical 
and ultraviolet filters ($W2$, $M2$, $W1$, $U$, $B$, $V$).
 The UVOT photometric system is described in \citet{key:poole2008}.
Photometry was computed from a 5\arcsec\ source region around 3C~279
using the publicly available UVOT {\tt ftools} data-reduction
suite.
The background region 
was taken from an annulus with inner and outer radii of $27\farcs5$ and $35\arcsec$, respectively.
Galactic extinction for each band in the direction of 3C 279 was adopted as given in Table~\ref{tab:Av}.

\begin{table}[bp]
\begin{center}
\caption{Galactic extinctions in the UV-optical-Near IR bands as used in this paper. The extinctions are based on the reddening of $E(B-V)=0.029$\,mag~\citep{Sch98} with $A_{V}/E(B-V)=3.1$. See also \citet{Lar08}.}
\label{tab:Av}
\small
\begin{tabular}{l l l}
 \hline  \hline
Band & $A_{\lambda}$ & Instruments  \\
\hline
$W2$ & 0.271 & UVOT\\
$M2$ & 0.285 & UVOT\\
$W1$ & 0.195 & UVOT\\
$U$  & 0.147 & UVOT\\
$B$ & 0.123 & UVOT, SMARTS\\
$V$ & 0.093 & UVOT, SMARTS, Kanata\\
$R, R_C$ & 0.075 & SMARTS, Kanata\\
$I_C$, & 0.056 & Kanata \\
$J$ & 0.027  & SMARTS \\
$K$ & 0.010 & SMARTS \\
\hline 
\end{tabular}
\end{center}
\end{table}

\subsubsection{SMARTS: Optical-Near Infrared bands}
The source has been monitored for several years in the 
optical and NIR bands ($B$, $V$, $R$, $J$, and $K$ bands)
under the SMARTS project\footnote{\url{http://www.astro.yale.edu/smarts/glast/}}, organized by Yale University.
Data reduction and analysis are described in \cite{Bon12,Cha12}, 
and the typical uncertainties for a bright source like 3C~279 are $1-2$\%.
The publicly available data were provided in magnitude scale.
In a similar manner as presented in \citet{Nal12},  
the data in magnitude scale ${m_\lambda}$ were converted into flux densities 
as $\nu F_{\nu}({\rm erg}\,{\rm s}^{-1}\,{\rm cm}^{-2})=10^{(Z_{\lambda} - m_{\lambda}+A_{\lambda})}$,
where $Z_{\lambda}$ is an effective zero point and $A_{\lambda}$ is the extinction for each band.
The effective zero point is calculated as
$Z_{\lambda} = 2.5\log{(c/\lambda_{eff} \times f_{\nu})}$, 
where $\lambda_{eff}$ and $f_{\nu}$ are parameters taken from 
Table A2 in \citet{Bes98}.
The extinctions were corrected using the values in Table~\ref{tab:Av}.

\subsubsection{The Kanata telescope: Optical photopolarimetry}

We performed the {\it V}, {\it R$_C$} and {\it I$_C$}-band photometry and
{\it R$_C$}-band polarimetry observations of 3C~279 using HOWPol instrument 
installed on the 1.5\,m Kanata telescope located at the 
Higashi-Hiroshima Observatory, Japan \citep{Kaw08}.
We obtained 36 daily photometric measurements in each band,
and 35 polarimetry measurements in the {\it R$_C$} band.

A sequence of photopolarimetric observations consisted of successive exposures 
at four position angles of a half-wave plate: $0^{\circ}$, $45^{\circ}$, $22.5^{\circ}$, and $67.5^{\circ}$.
The data were reduced using standard procedures for CCD photometry.
We performed aperture photometry using the \verb|APPHOT| package in \verb|PYRAF|
\footnote{PYRAF is a product of the Space Telescope Science Institute, which is operated by AURA for
NASA. See \url{http://www.stsci.edu/institute/software\_hardware/pyraf}},
and the differential photometry with a comparison star taken in the
same frame of 3C~279.
The comparison star is located at R.A. = 12:56:14.4 and Decl. = $-$05:46:47.6 (J2000),
and its magnitudes are  {\it V} = 15.92, {\it R$_C$}$ = 15.35$ \citep{Bon12}
and {\it I$_C$}$= 14.743$ \citep{Zac09}.
The data have been corrected for Galactic extinction 
as summarized in Table~\ref{tab:Av}.

Polarimetry with the HOWPol suffers from large instrumental polarization ($\delta$PD $\sim$ 4\%) 
caused by the reflection of the incident light on the tertiary mirror of the telescope.
The instrumental polarization was modeled as a function of the declination of the object and the
hour angle at the observation, and was subtracted from the observed value.
We confirmed that the accuracy of instrumental polarization subtraction was better than 0.5\%
in the {\it R$_C$} band using unpolarized standard stars. 
The polarization angle is defined as usual
(measured from north to east), 
based on calibrations with polarized stars, HD183143 and HD204827 \citep{Sch83}.
We also confirmed that the systematic error caused by instrumental 
polarization was smaller than $2^{\circ}$ using the polarized stars.

\subsection{Radio observations}
\subsubsection{SMA: millimeter-wave band} 

The 230\,GHz flux density data was obtained at the Submillimeter Array (SMA), an 8-element interferometer located near the summit of Mauna Kea (Hawaii). 3C~279 is included in an ongoing monitoring program at the SMA to determine the flux densities of compact extragalactic radio sources that can be used as calibrators at millimeter and sub-millimeter wavelengths~\citep{Gur07}. Observations of available potential calibrators are from time to time observed for 3 to 5 minutes, and the measured source signal strength calibrated against known standards, typically solar system objects (Titan, Uranus, Neptune, or Callisto). Data from this program are updated regularly and are available at the SMA website\footnote{\url{http://sma1.sma.hawaii.edu/callist/callist.html}}.

\section{Multi-band Observational Results}
\label{sec_multiband}

\subsection{Light curve}

The multi-band light curves from the $\gamma$-ray to the radio 
bands taken between MJD~56615 
and 56775, are shown in Figure~\ref{MWLLC} (covering 
the same period as in Figure~\ref{LC}).
The $\gamma$-ray light curve measured by \Fermi-LAT is plotted 
using one-day time bins.  The X-ray fluxes were measured 
by \Swift-XRT in the 0.5--5\,keV band.  The third panel shows fluxes 
in the optical $V$-band measured by \Swift-UVOT, SMARTS and Kanata
as well as the $R$-band data measured by SMARTS and Kanata.
The optical polarization data were measured by Kanata in the $R_C$-band.
The 230\,GHz fluxes were based on the results from 
SMA and also included results by ALMA\footnote{Taken from the ALMA Calibrator 
Catalogue, \url{https://almascience.eso.org/sc/}}.
In the plot, the periods (A--D) as defined in 
Table~\ref{tab:GammaSPfit} are also indicated.

Generally, the source showed the most active states in the $\gamma$-ray 
band at the beginning (including Period B, Flare 1) and the end 
(including Period D, Flare 3) of the epoch considered in this paper.
In the X-ray band, we also see two high-flux states, 
in the first half and in the second half of this epoch.  While in the 
first active phase the flux variation was not apparently well correlated 
between the $\gamma$-ray and the X-ray bands, we can see flaring 
activities in both the $\gamma$-ray the X-ray bands around 
Period D ($\sim$MJD~56750).

During the epoch considered here, the optical flux showed 
significantly different behavior than that in the $\gamma$-ray and the X-ray bands.
In the beginning of this epoch, the measured fluxes were 
relatively low with relatively high polarization degrees, 
of $\sim 20\%$. Around period B, the $\gamma$-ray 
showed a very rapid flare with a hard photon index, but the 
source did not show any enhanced optical fluxes.
After that, the optical fluxes started increasing gradually, 
with a drop of the polarization degree to $\sim5\%$ after Period C.
The $\gamma$-ray and X-ray band fluxes dropped,
but the optical flux still continued increasing, 
and peaked at $sim$MJD~56720. In the largest flaring event 
in Period D, where the $\gamma$-ray ($>100$\,MeV) and X-ray 
fluxes were highest, the optical 
flux showed only minor enhancement, and had already started 
decreasing from its peak value.  The optical polarization angle did 
not show any rotation throughout the observations considered here, and 
remained rather constant around $50^{\circ}$ with respect to 
the jet direction observed by Very Long Baseline Interferometry 
observations at radio bands \citep[e.g.,][]{Jor05}. 

The 230\,GHz flux was less variable compared to other bands, 
varying by about 50\%, from $\sim 8$\,Jy to $\sim 12$\,Jy.
Even though the amplitude of the variation was much smaller, the 
general variability pattern of the 230\,GHz band followed a similar 
pattern to that seen in the optical; a low state in the beginning of
the epoch, followed by increased activity in the middle, and a decrease 
toward to the end of the interval. No prominent millimeter-wave flares corresponding 
to the large $\gamma$-ray flaring events (Flares 1--3) were observed.

\subsection{Spectral energy distributions}

\begin{figure*}[htbp]
\centering
\plotone{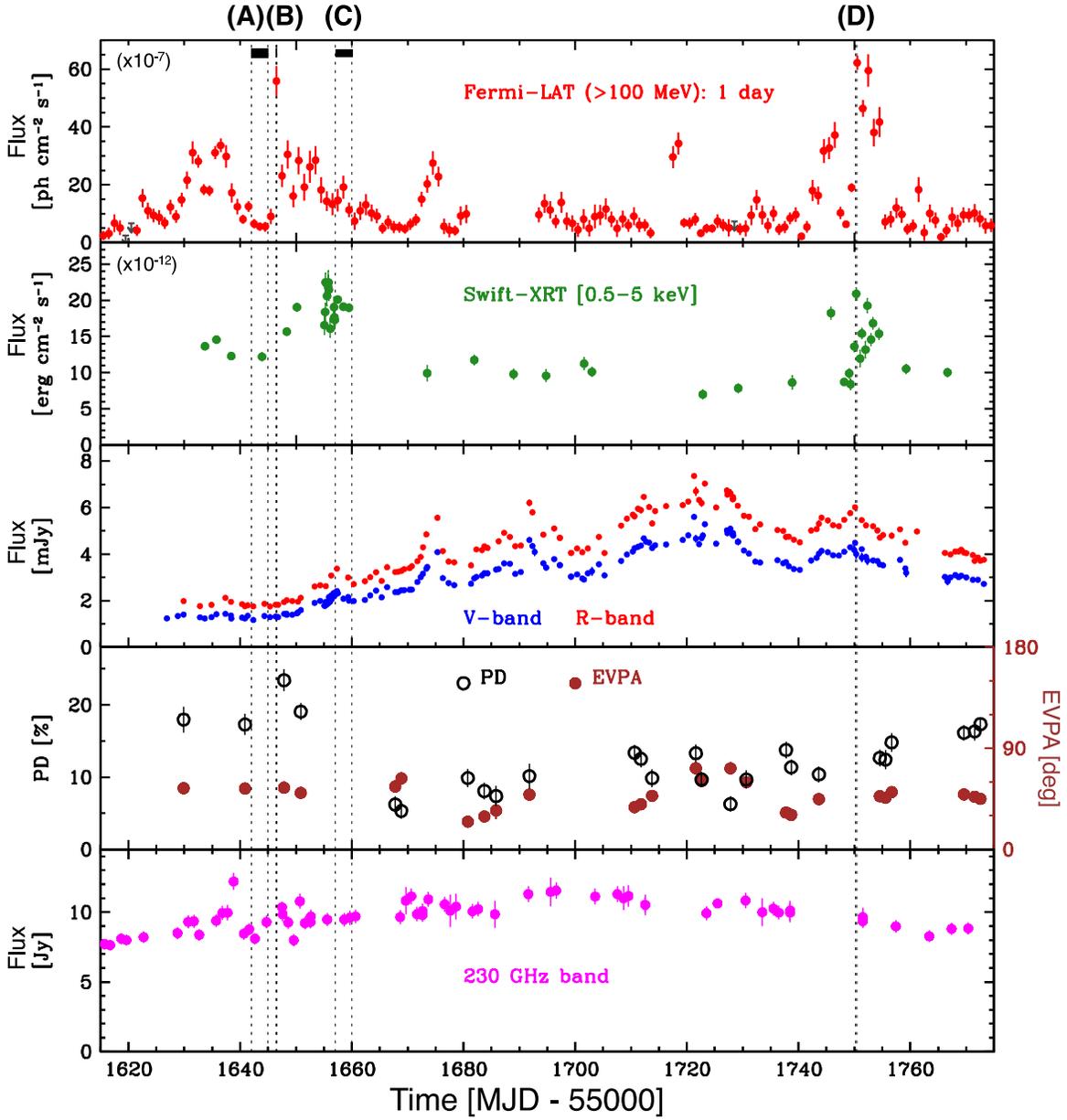}
\caption{Multiwavelength light curves of 3C~279 covering the same period as in Figure~\ref{LC}.
From the top, the panels show:
(1) $\gamma$-ray photon flux above 100\,MeV in 1-day bins from \Fermi-LAT;
(2) X-ray flux density between 0.5--5 keV from \Swift-XRT;
(3) optical flux density from \Swift-UVOT, SMARTS ($V,R$), and Kanata ($V$);
(4) optical polarization degree (scale on the left) and electric vector polarization angle (EVPA, scale on the right) from Kanata;
(5) mm flux density (230\,GHz) measured by SMA and ALMA.
The vertical dotted lines indicate the periods (A$-$D) as defined in Table~\ref{tab:GammaSPfit} when the $\gamma$-ray spectra were extracted.
A gap in the $\gamma$-ray data by \Fermi-LAT around MJD~56680--56690 is due to a ToO observation of the Crab Nebula, during which time no exposure was available in the direction of 3C~279.
The vertical bars in data points represent 1\,$\sigma$ statistical errors and 
the down arrows indicate 95\% confidence level upper limits.
}
\label{MWLLC}
\end{figure*}

\begin{figure*}[htbp]
\centering
\plotone{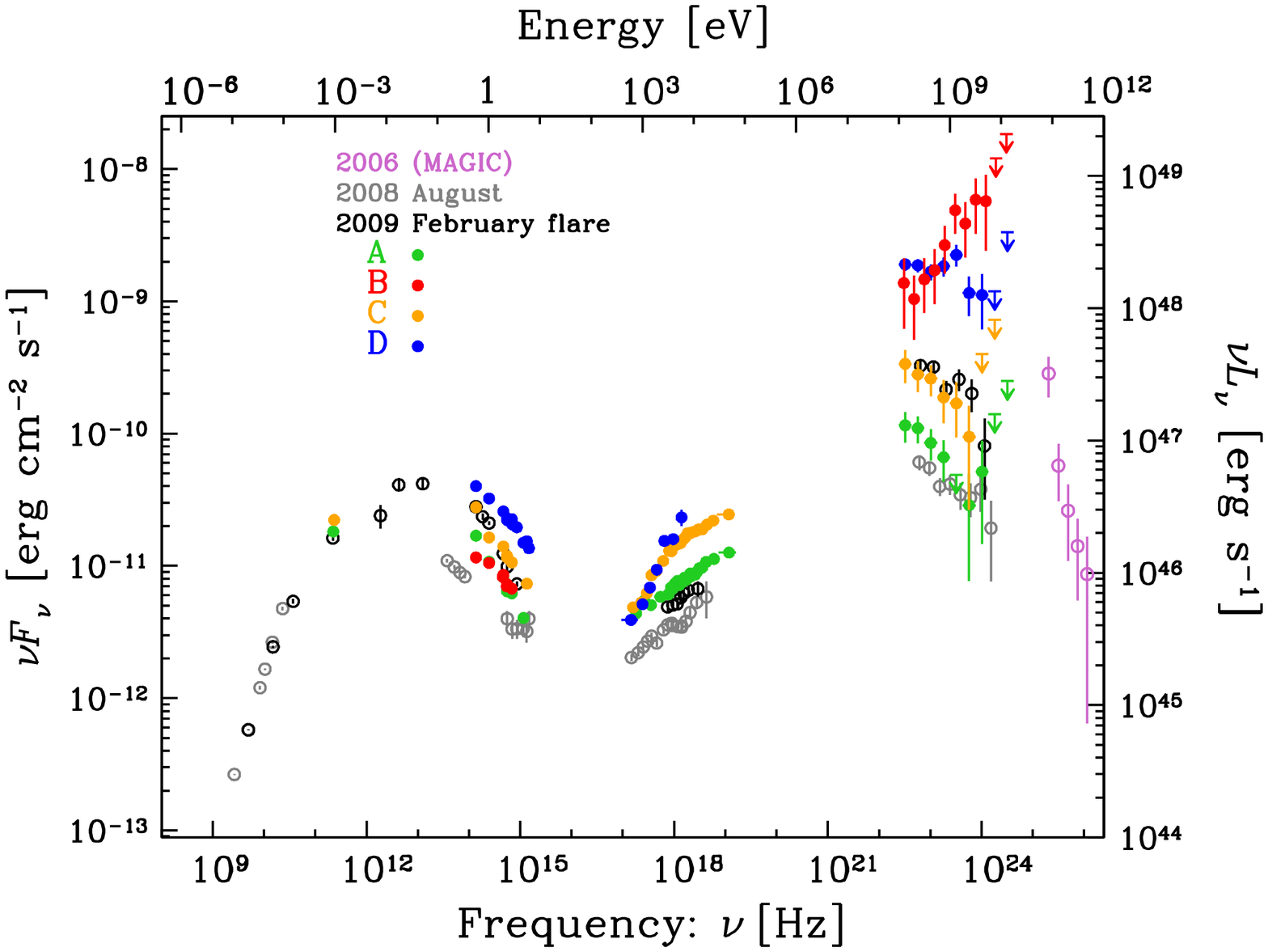}
\caption{Broadband spectral energy distributions of 3C\,279 for the four observational periods 
defined in Section~\ref{sec_fermi} (see also Table~\ref{tab:GammaSPfit} and Figure~\ref{MWLLC}).
The vertical bars in data points represent 1\,$\sigma$ statistical errors and 
the down arrows indicate 95\% confidence level upper limits.
The plot includes historical SEDs of 3C\,279 in a low state (in 2008 August) and in a flaring state (in 2009 February) from the 2008-2010 campaign \citep{hay12}. The measured spectral fluxes by MAGIC in 2006 are also plotted~\citep{MAGIC}.
}
\label{allSED}
\end{figure*}

Figure~\ref{allSED} shows broadband SEDs for each period as defined 
in Table~\ref{tab:GammaSPfit} (see also Figure~\ref{MWLLC} in the light curves).
The data sets include \Fermi-LAT (see also Figure~\ref{GammaSP}), 
\NuSTAR\ (for Periods A and C) and \Swift-XRT (for Periods A, C, and D), 
\Swift-UVOT ($W1$ for Period A, $M2$ 
and $V$ for Period C, all six bands for Period D), 
SMARTS ($B$, $V$, $R$, $J$, $K$ bands for all four periods) and Kanata 
($V$, $R_C$ bands for Period B), and SMA (for Periods A and C).
All data in the 
figure were taken within the time spans as defined in 
Table~\ref{tab:GammaSPfit}, but with a very slight offset in some 
optical data as follows: 
for Period B (MJD~56646.4--56646.6),
SMARTS data were taken during MJD 56646.348--56646.353  
and Kanata data were taken starting at MJD 56646.8078.
The SMARTS data observed during MJD 56750.1986--56750.2045 were for Period D (MJD 56750.210--56750.477). 
Unfortunately, no X-ray observation was performed during Period B. 
For comparison, the SEDs during a polarization change associated with a $\gamma$-ray flare observed in 2009 February, a low state in 2008 August~\citep{hay12}, and very-high-energy $\gamma$-ray spectral points measured by MAGIC in 2006~\citep{MAGIC} are also included.

\section{Discussion}
\label{sec_disc}

This campaign on blazar 3C~279 has three observational results of primary interest: 
(1) a rapid $\gamma$-ray flare with very hard $\gamma$-ray spectrum and no optical 
counterpart observed by \Fermi-LAT peaking at MJD~56646 (Flare 1); 
(2) intraday variability and a stable hard X-ray spectrum observed by \NuSTAR\, 
in combination with significant X-ray spectral variations observed by \Swift-XRT; and 
(3) a long trend of increasing optical flux without a corresponding increase 
in the $\gamma$-ray flux. These results are discussed in detail in the following subsections.

\subsection{Extreme $\gamma$-ray flare}
\label{sec_disc_gamma}

The $\gamma$-ray flares peaking at MJD~56646 (Flare 1, Period B) and MJD~56750 
(Flare 3, Period D) are the brightest flares detected in 3C~279 by the \Fermi-LAT.
With photon fluxes of $\simeq 1.2\times 10^{-5}$\,\phcms 
above $100\;{\rm MeV}$, they are brighter by a factor $\simeq 4$ than the flare 
peaking at MJD~54880 \citep{Nature}, and comparable to the record fluxes detected
by EGRET \citep{Weh98}.

Flare 1 was characterized by unprecedented rapid  
variability, with a flux-doubling time scale estimated conservatively
at $t_{\rm var} \simeq 2\;{\rm h}$, and a very hard $\gamma$-ray spectrum, with
photon index $\Gamma_\gamma \simeq 1.7$. 
Notably, we do not detect any simultaneous activity 
in the optical band. 
These facts make Flare 1 different from any previous 
$\gamma$-ray activity observed in 3C~279. The closest analog was a rapid $\gamma$-ray flare 
in PKS~1510$-$089 peaking at MJD~55854 with a flux-doubling time scale of $\simeq 1\;{\rm h}$ 
and photon index of $\Gamma_\gamma \simeq 2$ \citep{2013ApJ...766L..11S,2013MNRAS.430.1324N}. 
However, in that case there were no simultaneous multiwavelength observations 
because of the proximity of PKS~1510$-$089 to the Sun.

\begin{figure*}[tb]
\centering
\includegraphics[width=2.06\columnwidth]{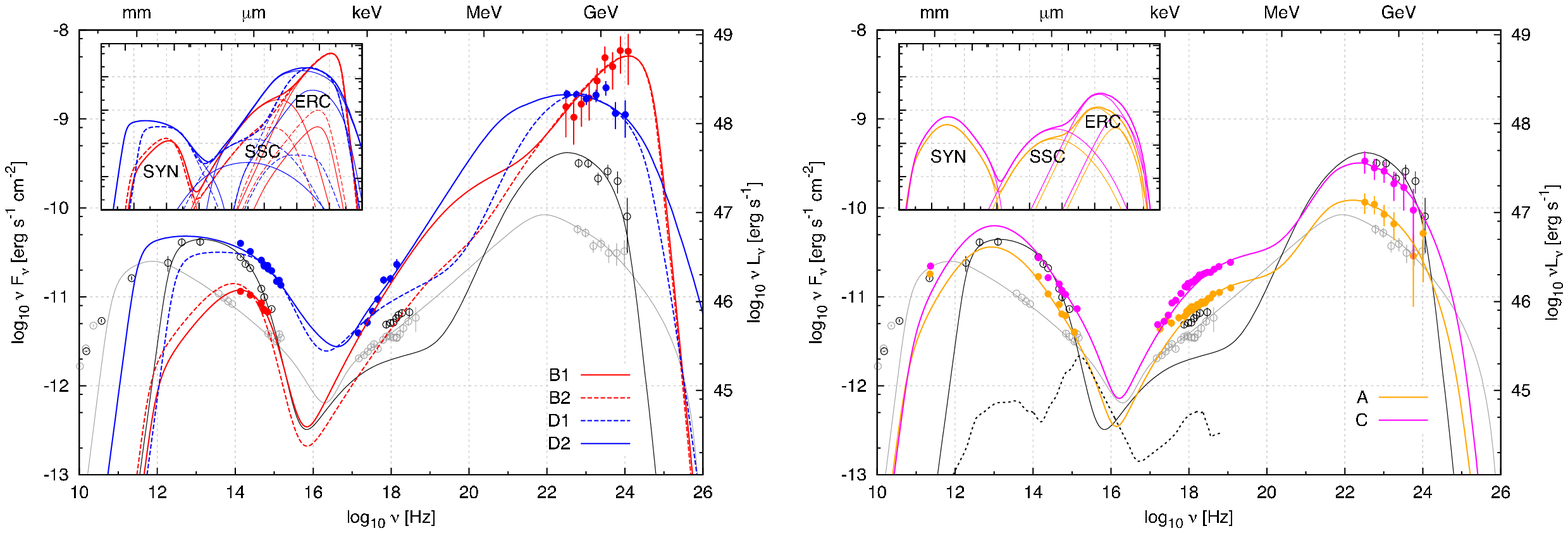}
\caption{{\it Left panel}: Spectral energy distributions of 3C~279 during the 
brightest $\gamma$-ray flares --- Flare 1 (Period B, red points) 
and Flare 3 (Period D, blue points) --- see Section \ref{sec_disc_gamma} for discussion. 
{\it Right panel}: SEDs during two \NuSTAR\ pointings --- Period A 
(orange points) and Period C (magenta points) --- see 
Section \ref{sec_disc_Xray} for discussion.
Solid and dashed lines show SED models obtained with the leptonic code {\tt Blazar}. 
Model parameters are listed in Table \ref{tab_sed_param}.
Black and gray lines show historical data and SED models from \cite{hay12}.
Black dashed line shows the composite SED for radio-loud quasars \citep{Elv94} normalized to $L_{\rm d} = 6\times 10^{45}\;{\rm erg\,s^{-1}}$.
The inset illustrates schematically the decomposition of each SED 
model into contributions from individual radiative mechanisms: 
(in order of increasing peak frequency) synchrotron, SSC, ERC(IR), 
ERC(BLR).
The axes and line types are the same as in the main plot.
}
\label{fig_sed}
\end{figure*}

The SED for Flare 1 is presented in Figure~\ref{allSED} and \ref{fig_sed} as Period B. The very hard $\gamma$-ray 
spectrum measured by \Fermi-LAT requires that the high-energy SED peak lies at 
energy $> 2\;{\rm GeV}$. We assume conservatively the actual SED peak lies 
at $\simeq 5\;{\rm GeV}$, so that the apparent $\gamma$-ray peak $\nu L_\nu$ 
luminosity is $L_\gamma \simeq 6\times 10^{48}\;{\rm erg\,s^{-1}}$. We further 
assume that this $\gamma$-ray emission is produced by the external radiation 
Comptonization (ERC) mechanism (see \citealt{2009ApJ...704...38S} for a review 
of alternative mechanisms). The corresponding synchrotron component is expected 
to peak close to the optical band. The observed 
simultaneous optical/UV spectrum, and the lack of simultaneous optical variability, 
place very strong constraints on the Compton dominance 
parameter $q = L_\gamma / L_{\rm syn} \gtrsim 300$.

\begin{figure*}[htp]
\centering
\includegraphics[width=1.35\columnwidth]{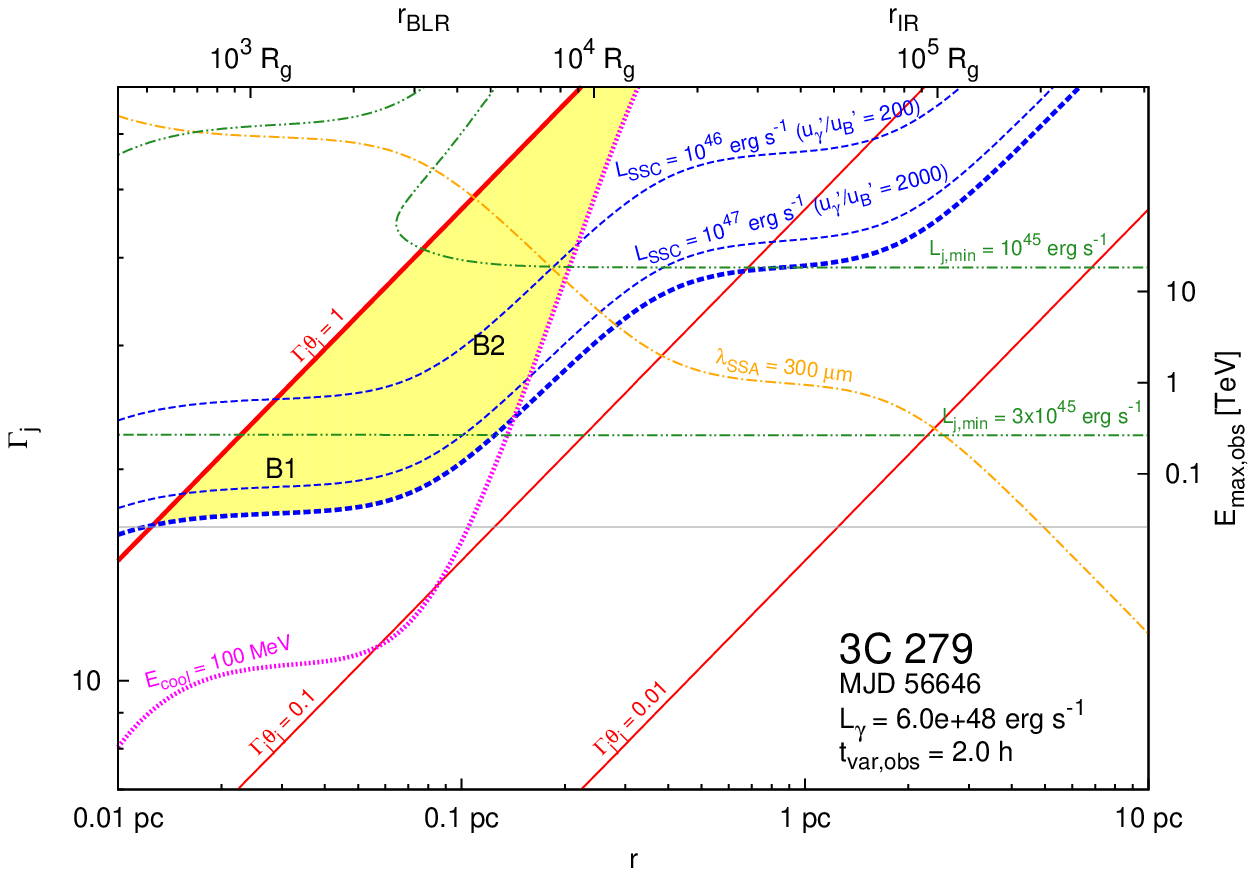}
\caption{Constraints on the parameter space of location $r$ and Lorentz factor $\Gamma_{\rm j}$ for the emitting region producing $\gamma$-ray Flare 1 (Period B) \citep[see][for detailed description of the model]{2014ApJ...789..161N}.
The following constraints are shown: from jet collimation parameter $\Gamma_{\rm j}\theta_{\rm j}$ (solid red lines), from SSC luminosity $L_{\rm SSC}$ (dashed blue lines), from energy threshold for efficient cooling $E_{\rm cool}$ (dotted magenta line), from the characteristic wavelength of synchrotron self-absorption $\lambda_{\rm SSA}$ (dot-dashed orange line), from the minimum required jet power $L_{\rm j,min}$ (double-dot-dashed green lines), and from the characteristic energy of intrinsic $\gamma$-ray absorption $E_{\rm max,obs}$ (solid gray line). The region allowed by the collimation constraint ($\Gamma_{\rm j}\theta_{\rm j} < 1$), the SSC constraint ($L_{\rm SSC} < 2\times10^{47}\,{\rm erg}\,{\rm s^{-1}}$) and the cooling constraint ($E_{\rm cool} < 100\,{\rm MeV}$) is shaded in yellow. Two particular solutions B1 and B2 for which SED models are shown in Figure~\ref{fig_sed} are indicated.}
\label{fig_scal}
\end{figure*}

In order to constrain the location along the jet $r$ and the Lorentz factor $\Gamma_{\rm j}$ 
of the emitting region that produced Flare 1, we use the recent model 
of \cite{2014ApJ...789..161N}.
In that model, the allowed parameter space for the $\gamma$-ray emitting region is defined by three constraints: (1) a jet collimation constraint $\Gamma_{\rm j}\theta_{\rm j} < 1$, where $\theta_{\rm j}$ is the jet half-opening angle, (2) a constraint on the SSC luminosity $L_{\rm SSC} < L_{\rm X}$, where $L_{\rm X}$ is the observed X-ray, hard X-ray or soft $\gamma$-ray luminosity (depending 
on the expected energy of the SSC peak), and (3) a constraint on the radiative cooling time scale $E_{\rm cool} < 100\,{\rm MeV}$, where $E_{\rm cool}$ is the characteristic observed energy of the $\gamma$-ray photons produced by the electrons for which the radiative cooling time scale is comparable to the variability time scale.
The proper size of the emission region is estimated directly from the observed variability time scale $R \simeq \mathcal{D}ct_{\rm var,obs}/(1+z)$, and it is related to the jet opening angle by $R/r = \theta_{\rm j}$. 
In addition to the parameters discussed above, 
we adopt a standard ratio of the Doppler-to-Lorentz factors $\mathcal{D}/\Gamma_{\rm j} = 1$, 
and we also need to specify the upper limit on the expected synchrotron self-Compton 
(SSC) component. As the SSC component likely peaks in the MeV band, where we do not 
have any observational constraints, we will conservatively assume that 
$L_{\rm SSC} < L_\gamma/30 \simeq 2\times 10^{47}\;{\rm erg\,s^{-1}}$. The constraints 
on the parameter space are shown in Figure~\ref{fig_scal}. The yellow-shaded area indicates the allowed location of the $\gamma$-ray emitting region. 
For reasonable values of the jet Lorentz factor $20 < \Gamma_{\rm j} < 30$,
it should be located between $0.015\;{\rm pc} < r < 0.15\;{\rm pc}$,
where the external radiation is dominated by the broad emission lines ($r_{\rm BLR} \simeq 0.8\;{\rm pc}$).
At larger distances, the energy density of external radiation fields will be insufficient to provide efficient cooling of the electrons producing the $100\;{\rm MeV}$ photons on the observed time scales, and also it will be more difficult to maintain a sufficiently high energy density to power such a luminous $\gamma$-ray flare from a very compact emission region.
The Lorentz factor should be $\Gamma_{\rm j} > 17$, 
although this constraint would be stronger if we assumed a lower $L_{\rm SSC}$.
For the subsequent modeling of the SED, we will adopt two possible solutions indicated 
in Figure \ref{fig_scal}: (B1) $r = 0.03\;{\rm pc}$ and $\Gamma_{\rm j} = 20$, and (B2) $r = 0.12\;{\rm pc}$ and $\Gamma_{\rm j} = 30$. This corresponds to the jet opening angle $\theta_{\rm j}$ satisfying: (B1) $\Gamma_{\rm j}\theta_{\rm j} = 0.61$ and (B2) $\Gamma_{\rm j}\theta_{\rm j} = 0.34$.

We use the {\tt Blazar} code \citep{2003A&A...406..855M} to model the SED of 3C~279 during 
Flare 1 with a standard leptonic model including the synchrotron, SSC and ERC processes. 
The distribution of external radiation is scaled to the accretion disk luminosity of $L_{\rm d} \simeq 6\times 10^{45}\;{\rm erg\,s^{-1}}$ using 
standard relations for the characteristic radii of the broad-line region (BLR) and the dusty torus \citep{2009ApJ...704...38S} 
with covering factors $\xi_{\rm BLR} = \xi_{\rm IR} = 0.1$.
In general, we use a double-broken power-law energy distribution of injected electrons $N(\gamma) \propto \gamma^{-p_i}$ with two breaks at $\gamma_1$ and $\gamma_2$ and three indices: $p_1$ for $\gamma < \gamma_1$, $p_2$ for $\gamma_1 < \gamma < \gamma_2$, and $p_3$ for $\gamma > \gamma_2$.
In order to reproduce a very hard $\gamma$-ray spectrum in the fast-cooling regime, we set $p_1 = 1$.
The goal of the modeling is to match the ERC(BLR) peak with the observed $\gamma$-ray peak 
by adjusting the maximum electron Lorentz factor, $\gamma_1$, and the electron 
distribution normalization, and then to match the synchrotron component with the 
optical/UV data by adjusting the co-moving magnetic field $B'$. One should remember 
that because of the lack of any optical activity simultaneous with Flare 1, the 
synchrotron component should actually be below the optical/UV data points, so $B'$ 
should be treated more like an upper limit than an actual value.
The results of the 
SED modeling are presented in Figure \ref{fig_sed}, and essential model parameters are listed in Table \ref{tab_sed_param}.
For solution B1 we obtain $\gamma_1 = 3700$ and $B' = 0.31\;{\rm G}$,
and for solution B2 we obtain $\gamma_1 = 2800$ and $B' = 0.3\;{\rm G}$.

We consider the basic energetic requirements for producing such an SED. 
We can estimate the total required jet power as $L_{\rm j} \simeq L_\gamma/(\eta_{\rm j}\Gamma_{\rm j}^2)$, 
where $\eta_{\rm j} \sim 0.1$ is the radiative efficiency of the jet. 
And we can use the estimated magnetic field strength to calculate the magnetic 
jet power $L_{\rm B} = \pi R^2\Gamma_{\rm j}^2u_{\rm B}'c = (c/8)(\Gamma_{\rm j}\theta_{\rm j})^2(B'r)^2$, 
where $R = \theta_{\rm j}r$ is the jet radius, $\theta_{\rm j}$ is the jet 
half-opening angle, and $u_{\rm B}' = B'^2/(8\pi)$ is the magnetic energy density.
For solution B1 we obtain $L_{\rm j} \simeq 1.5\times 10^{47}\;{\rm erg\,s^{-1}}$ and $L_{\rm B} \simeq 1.1\times 10^{42}\;{\rm erg\,s^{-1}}$, and for solution B2 we obtain $L_{\rm j} \simeq 7\times 10^{46}\;{\rm erg\,s^{-1}}$ and $L_{\rm B} \simeq 5\times 10^{42}\;{\rm erg\,s^{-1}}$.
In both cases, 
the required magnetic jet power is a tiny fraction of the total 
jet power. This fraction is higher for solution B2, with 
$L_{\rm B}/L_{\rm j} \simeq 0.7\times10^{-4}$. This indicates that the 
emitting region responsible for Flare 1 is very strongly matter-dominated \citep[cf.][]{Jan14}, 
although several observed SEDs of 3C~279 in 2008-2010 can be described by an equipartition model~\citep{Der14}.

The total required jet power $L_{\rm j}$ appears very high compared to the accretion disk luminosity $L_{\rm d}$. For solution B1 we obtain $L_{\rm j}/L_{\rm d} \simeq 25$, and for solution B2 we obtain $L_{\rm j}/L_{\rm d} \simeq 11$. We note that there is no signature of increased disk luminosity in the UV data even for the lowest-flux state (Period A), which gives $L_{\rm d} < 9\times 10^{45}\;{\rm erg\,s^{-1}}$. 
Assuming the black hole mass of $M_{\rm bh} \simeq 5\times 10^8 M_\odot$ (see Section \ref{sec_intro}), the Eddington luminosity is $L_{\rm Edd} \simeq 8\times 10^{46}\;{\rm erg\,s^{-1}}$, hence $L_{\rm d}/L_{\rm Edd} \simeq 0.08$ and $L_{\rm j}/L_{\rm Edd} \simeq 0.9$ for solution B2. Solution B1 with moderate jet Lorentz factor $\Gamma_{\rm j} = 20$ predicts a super-Eddington jet power. The predicted jet power can also be decreased when assuming a higher jet radiative efficiency $\eta_{\rm j} > 0.1$.
Taking this into account, the energetic requirements for this flare are consistent with the typical relation between $L_{\rm j}$ and $L_{\rm d}$ for FSRQs \citep{Ghi14}.

In the standard scenario of magnetic acceleration of relativistic 
black-hole jets, the jets are initially dominated by the magnetic 
energy, which is gradually converted into the kinetic energy of 
plasma until the jet magnetization $\sigma \simeq L_{\rm B}/(L_{\rm j}-L_{\rm B}) \sim 1$ 
\citep{1994ApJ...426..269B,2007MNRAS.380...51K}. Additional conversion 
of the jet magnetic energy is possible locally under special conditions, e.g. 
rarefaction acceleration induced by rapid decline in external 
pressure \citep{2010NewA...15..749T,2010MNRAS.407...17K,2014PhPl...21g2124S}.
However, this additional jet acceleration is likely to lead to the loss of 
causal contact across the jet, and very wide opening angles 
($\Gamma_{\rm j}\theta_{\rm j} > 1$), more typical of gamma-ray bursts.
Magnetic energy can be dissipated directly in the process 
of magnetic reconnection.
However, it is unlikely that relativistic reconnection 
can operate with sufficient efficiency to convert highly 
magnetized plasma with $\sigma \sim 1$ to very weakly magnetized 
plasma with $\sigma \sim 10^{-4}$.
Additional matter could be injected into the jet by individual 
massive stars crossing the jet \citep{2013ApJ...774..113K},
or at the jet 
base during a brief pause in the jet production \citep{2014MNRAS.440.2185D}. 
The latter possibility potentially allows for a more uniform 
distribution of an unmagnetized plasma layer across the jet.

The very hard electron energy distribution with $p_1 \simeq 1$ extending up to $\gamma \gtrsim 2000$, 
required to explain the very hard $\gamma$-ray spectrum of Flare 1,
is challenging for many particle acceleration mechanism and emission 
scenarios \citep[\eg,][]{Bla95}.
Very hard electron spectra can be obtained in relativistic magnetic 
reconnection, but they require extremely high electron magnetization 
$\sigma_e > 100$ \citep{2014ApJ...783L..21S,2014arXiv1405.4040G,Wer14}. 
In the case of an electron-proton jet composition with no $e^+e^-$ pairs, 
one has $\sigma_e \simeq (\bar\gamma_p/\bar\gamma_e)(m_p/m_e) \sigma$, 
where $\bar\gamma_p$ and $\bar\gamma_e$ are typical Lorentz factors 
of protons and electrons, respectively. In principle, it is possible 
that $\sigma_e \gg \sigma$, so that such extreme acceleration of 
electrons is possible even in the case of $\sigma \lesssim 1$. 
The final outcome of the acceleration depends on how the dissipated 
magnetic energy is shared between the protons and electrons; the first 
study of relativistic electron-ion reconnection suggests roughly 
equal energy division \citep{2014AAinpress}.

We propose that Flare 1 of 3C~279, together with the similar flare of 
PKS~1510$-$089 peaking at MJD~55854 \citep{2013ApJ...766L..11S}, 
constitute an emerging class of rapid $\gamma$-ray events characterized 
by flux-doubling time scales of a couple of hours, very hard $\gamma$-ray 
spectra with spectral peaks in the GeV band, and significant time 
asymmetry with longer decay time scales \citep{2013MNRAS.430.1324N}. 
Moreover, the results of this work indicate that such events do not 
have significant multiwavelength counterparts. Since only two clear 
examples were identified in bright blazars during $\sim 6$ years so far of the {\it Fermi} 
mission, they appear to be rare events, and may not represent typical conditions 
of dissipation and particle acceleration in blazar jets.

\begin{table}[htp]
\begin{center}
\caption{Parameters of the SED models presented in Fig. \ref{fig_sed}.}
\label{tab_sed_param}
%\vskip 1em
\small
\begin{tabular}{lllllll}
\hline \hline
Model & A & B1 & B2 & C & D1 & D2 \\
\hline 
$r\;{\rm [pc]}$   & 1.1  & 0.03 & 0.12 & 1.1  & 0.03 & 1.1  \\
$\Gamma_{\rm j}$  & 8.5  & 20   & 30   & 10.5 & 25   & 30   \\
$\Gamma_{\rm j}\theta_{\rm j}$
                  & 1    & 0.61 & 0.34 & 1    & 1    & 1    \\
$B'\;{\rm [G]}$   & 0.13 & 0.31 & 0.3 & 0.13 & 1.75 & 0.14 \\
$p_1$             & 1    & 1    & 1    & 1    & 1    & 1.6  \\
$\gamma_1$        & 1000 & 3700 & 2800 & 1000 & 200  & 100  \\
$p_2$             & 2.4  & 7    & 7    & 2.4  & 2.5  & 2.5  \\
$\gamma_2$        & 3000 & ---  & ---  & 3000 & 2000 & 6000 \\
$p_3$             & 3.5  & ---  & ---  & 3.5  & 5    & 4    \\
\hline 
\end{tabular}
\end{center}
\end{table}

In Figure \ref{fig_sed}, we also present two SED models for Flare 3 (Period D). This flare is characterized by a typical $\gamma$-ray spectrum, and a more typical Compton dominance, as compared to Flare 1. In addition, for Flare 3 we have simultaneous UV and X-ray data from \Swift. The soft X-ray spectrum is very hard, with $\Gamma_{\rm X} = 1.22 \pm 0.07$.
We first attempted --- in model D1 located in the BLR --- to explain this X-ray spectrum by SSC emission from a very hard electron energy distribution ($p_1 = 1$). By coincidence, model B1 described in the previous subsection does exactly that. However, since the $\gamma$-ray spectrum for Period D is much softer than the exceptionally hard $\gamma$-ray spectrum for Period B, in model D1 we need to adopt a break in the electron energy distribution at $\gamma_{\rm br} \lesssim 200$, which shifts the observed peak of the SSC component to $\sim 100\;{\rm keV}$; hence the X-ray part of the SSC spectrum is too soft to explain the observed X-ray spectrum. 
We note that \cite{Pal15} present an SED model for a period overlapping with our Period D, in which the X-ray spectrum is matched with the SSC component. They made the model by adopting a higher value of $\gamma_{\rm br}$, which requires a superposition of ERC(BLR) and ERC(IR) components to explain the $\gamma$-ray spectrum.
In model D2 we were able to explain the X-ray spectrum with the low-energy tail of the ERC emission. The emission region in model D2 is located outside the BLR, and the entire high-energy component is strongly dominated by the ERC(IR) emission. We adopted a higher jet Lorentz factor $\Gamma_{\rm j} = 30$ and the low-energy electron distribution index $p_1 = 1.6$ for $\gamma < 100$ (see Table \ref{tab_sed_param}). While model D2 matches the observed X-ray spectrum, because it is located at relatively large distance $r \simeq 1.1\;{\rm pc}$, it predicts a rather long observed variability time scale of $t_{\rm var,obs} \sim 2\;{\rm d}$.

Such extremely high flux spectra as observed in Periods B and D could in some cases extend to even higher energies, and may possibly be detectable by ground-based Cherenkov telescopes (MAGIC, H.E.S.S., VERITAS, CTA).
Despite its moderate redshift, 3C 279 was detected twice
by MAGIC \citep{MAGIC,2011A&A...530A...4A} before the {\it Fermi} era. 
In this work, we have argued that the $\gamma$-rays originate at a radius $\sim 0.1\;{\rm pc}$, which is comparable with the estimated size of the broad emission line region $r_{\rm BLR}$ based upon reverberation mapping campaigns of other AGN \citep[e.g.,][]{Ben06,Kas07}.
This radius is also roughly comparable to the minimum radius from which the highest energy photon observed during our campaign --- $E_{\rm obs} = 26.1\;{\rm GeV}$ ($E \sim 40\;{\rm GeV}$ in the quasar rest frame) in Period A --- can escape without pair production. 
The highest energy photons detected in Periods B and D were $10.4\;{\rm GeV}$ and $13.5\;{\rm GeV}$, respectively (see Figure~\ref{LC}).
We could not distinguish if the non-detection of $\gtrsim 15\;{\rm GeV}$ photons was due to the absorption by the BLR photons or just the statistical limitation of the short integration time for the spectra.
Our emission models indicate a sharp drop in the source intrinsic spectral shape at $>10\;{\rm GeV}$ energies due to adopting a very steep high-energy electron distribution index $p_2 = 7$. In addition, the $\gamma$-ray emission above $10\;{\rm GeV}$ produced in the ERC(BLR) process is suppressed due to the reduction of the scattering cross section in the Klein-Nishina regime.

The importance of $\gamma$-ray absorption in the pair production process depends on the abundance of soft photons produced in the jet environment.
One source of soft photons is the emission lines radiated by the broad emission line clouds. 
The optical depth depends on the geometrical shape of the BLR, and is significantly reduced for the flat geometries \citep{Tav12,Ste14}.
Specifically, the results of \cite{Tav12} calculated for
$L_{\rm d} = 5\times 10^{45}\;{\rm erg\,s^{-1}}$ and for an intermediate geometric case indicate that absorption from BLR photons is not significant for $\gamma$-ray photons with $E_{\rm obs} \lesssim 50\;{\rm GeV}$ emitted at $r \sim r_{\rm BLR}$, and those with $E_{\rm obs} \lesssim 20\;{\rm GeV}$ at $r \sim 0.1r_{\rm BLR}$. 
However, other models of the BLR can be expected to set a larger lower bound on the emission distance scale $r$.
Another source of soft photons is Thomson scattering by the hot inter-cloud medium that is commonly invoked to confine the clouds by ram pressure.
If we adopt the spectral component associated with the accretion disk by \cite{1999ApJ...521..112P}, then the radius of the `$\gamma$-sphere' (for the $\sim 40\;{\rm GeV}$ photons in the source frame) is $\sim 0.2 (\left<\tau_T\right>/0.01)\;{\rm pc}$, where $\left<\tau_T\right>$ is the mean Thomson depth \citep[e.g.,][]{Bla95}.

The difficulties faced by the leptonic models do not necessarily mean that they should be abandoned, unless there exists a better alternative. Hadronic models have been applied to 3C~279 \citep{Boe09,Pet12,Dil15}, however, they always require an extremely large jet power of order $10^{49}\;{\rm erg\,s^{-1}}$, which is difficult to reconcile with observations of radio galaxies and theories of jet launching \citep{Zdz15}.

\subsection{Spectral variability in X-ray and hard X-ray bands}
\label{sec_disc_Xray}

The X-ray and hard X-ray behavior of 3C~279 revealed by \Swift-XRT and \NuSTAR\
appears to be quite complex.
This is especially interesting since the origin of 
this emission in FSRQ-type blazars is in general controversial and poorly 
understood \citep{2013ApJ...779...68S}. Here, we focus on constraining the 
mechanism of X-ray and hard X-ray emission observed during the two 
\NuSTAR\ pointings (Periods A and C).

The X-ray flux observed during the \NuSTAR\ pointings is relatively high 
for 3C~279, higher by a factor of $\sim 5$ than typical X-ray fluxes measured 
during the 2008--2010 campaign \citep{hay12}.
In Period C, the X-ray flux is higher by a factor of $\sim 2$ than in Period A, 
and it also shows some intraday variations. A comparison of the broadband SEDs 
for Period A and C suggests a roughly linear relation between optical, 
X-ray and $\gamma$-ray variations. However, this may be misleading, as the 
optical flux shows a long-term systematic increase that is not evident 
in the X-ray and $\gamma$-ray light curves. 
With typical photon index $\Gamma_{\rm X} \simeq 1.7$, 
the observed X-ray spectrum is relatively hard, with a spectral break 
observed in Period C at $\simeq 3.5\;{\rm keV}$. 
The \Swift-XRT data show a clear anti-correlation between the soft X-ray 
photon index and the soft X-ray flux (see Figure~\ref{SwiftXray}).

The X-ray emission of FSRQs is typically attributed either to the SSC 
emission of medium-energy electrons, or to the ERC emission of low-energy 
electrons. 
Adopting a one-zone model in the SSC scenario for the X-ray band,
we attempted to explain the observed broadband SEDs for Periods A and C
together with the synchrotron emission for the optical band
and the ERC component for the $\gamma$-ray band.
Figure~\ref{fig_sed} shows the SEDs for Periods A and C with the emission models
based on the parameters in Table~\ref{tab_sed_param}.
Actually, the observed high X-ray flux and hard X-ray spectra challenge the SSC scenario.
In order to match the relatively high observed X-ray 
luminosity with that expected to be produced via 
SSC, as well as the corresponding synchrotron component 
and the ERC components with a condition of $\Gamma_{\rm j}\theta_{\rm j} = 1$,
one needs to adopt a rather low jet Lorentz factor $\Gamma_{\rm j} \simeq 8.5$ for Period A
and $\Gamma_{\rm j} \simeq 10.5$ for Period C.
A low jet Lorentz factor requires a more powerful jet; in 
the case of Period C with $\gamma$-ray luminosity 
$L_\gamma \simeq 4\times 10^{47}\;{\rm erg\,s^{-1}}$ we estimate 
$L_{\rm j} \simeq L_\gamma/(\eta_{\rm j}\Gamma_{\rm j}^2) 
\simeq 4\times 10^{46}(\eta_{\rm j}/0.1)^{-1}\;{\rm erg\,s^{-1}}$. The jet 
Lorentz factor can be higher, and the required jet power lower, by 
allowing that $\Gamma_{\rm j}\theta_{\rm j} < 1$. 
Typical X-ray SEDs 
from the SSC component in FSRQs are flat; it is possible to obtain 
a hard SSC spectrum by choosing an electron energy distribution 
peaking at $\gamma_{\rm peak} \gtrsim 300$. 
However, in order to reproduce 
apparent linear relations between 
optical, X-ray and $\gamma$-ray fluxes from Periods A to C,
it was necessary to adjust the value of $\Gamma_{\rm j}$ to compensate for the quadratic 
dependence of the SSC luminosity on the synchrotron luminosity 
$L_{\rm SSC} \propto L_{\rm syn}^2$. Therefore, while it is possible to make 
a one-zone model that fits the observed SEDs for both Periods A and C with 
X-rays produced by the SSC process and with reasonable jet power, 
such a model cannot naturally account for the observed flux variations 
over the multiwavelength bands.
We note that lower Lorentz factors $\Gamma_{\rm j} \simeq 10$ required for modeling the SEDs for 
Periods A and C, together higher Lorentz factors $\Gamma_{\rm j} \simeq 30$ 
required for modeling the SEDs for Periods B and D, could indicate the 
existence of a spine-sheath jet structure \citep{Ghi05}. In such case 
the $\gamma$-ray flares would be produced in the fast spine and the 
bulk of X-ray and hard X-ray emission in the slow sheath.

In the ERC scenario, on the other hand, the observed X-ray flux could be dominated by the 
low-energy tail of either the ERC(BLR) or ERC(IR) components, depending 
on the location of the emitting region. In one-zone models, this would 
require the X-ray spectra to be related to the $\gamma$-ray spectra 
by a single spectral component. This would explain the apparent linear 
relation between the X-ray and $\gamma$-ray data, although there are 
insufficient X-ray observations in the current campaign to probe the 
direct correlation between X-ray and $\gamma$-ray fluxes. (During the 
previous campaign on 3C~279, no significant correlation was detected 
between the X-ray and $\gamma$-ray fluxes; \citealt{hay12}). Judging from 
the simultaneous X-ray, hard X-ray and $\gamma$-ray SEDs, it could be 
possible to connect them by a single spectral component, especially for 
Period A. This would require a power-law extension of the X-ray spectrum 
all the way to the low-energy end of the $\gamma$-ray spectrum. This is possible 
only for the ERC(IR) component where no cooling break is expected. However, 
because the low-energy ERC(IR) emission would be produced deep in the 
slow-cooling regime, very little flux variability would be expected on 
daily time scales. The ERC(BLR) component is very likely to feature a 
cooling break, and possibly an additional low-energy spectral break 
at $2.6(\Gamma_{\rm j}/20)^2\;{\rm keV}$ produced by trans-relativistic 
electrons ($\gamma \sim 1$).

We conclude that there is no one-zone leptonic SED model that can satisfactorily 
explain the production of X-ray emission observed by \NuSTAR\ in 3C~279. Various 
alternative mechanisms can be proposed where the observed X-ray emission is 
produced at a different location from the optical and/or $\gamma$-ray emission. 
Observation of a transient spectral break at $\simeq 3.5\;{\rm keV}$ in Period C may 
indicate a superposition of two spectral components in the X-ray band.
Similar spectral breaks observed in high-redshift FSRQs were interpreted 
as due to very strong absorption \citep{2001MNRAS.323..373F}.  However in our 
case this interpretation is challenged by the lack of a break in Period A, only two weeks earlier.
Alternative mechanisms for the X-ray emission include IC emission from the 
accretion-disk corona, ERC or synchrotron emission from the jet acceleration region, and 
hadronic mechanisms \citep{Boe09}. 
However, for most mechanisms it may be challenging to explain the relatively 
high X-ray luminosity $L_{\rm X} \simeq 2\times 10^{46}\;{\rm erg\,s^{-1}}$ 
observed during Period C, a factor $\sim 3$ higher than $L_{\rm d}$.
This certainly excludes the accretion-disk coronal emission proposed 
tentatively for PKS~1510$-$089 in the low state \citep{Nal12}. The `jet base' 
scenario, where luminous weakly beamed X-ray emission is produced at very short
distances from the supermassive black hole at which the jet is not yet fully 
accelerated, is motivated by recent observations of misaligned 
FSRQs \citep{2014ApJ...791..119B}, and will be investigated in detail elsewhere.

\subsection{Optical behavior}
\label{sec_disc_optical}

The observed optical flux shows a systematic increase by a factor of $\sim 4$ over 
a period of $\sim 80$ days (MJD 56645--56725). Superposed on this trend 
are weak flares that correspond very poorly to the strong $\gamma$-ray flares. 
No similar systematic trend is seen in the X-ray and $\gamma$-ray light curves.
This is in contrast with the good overall correlation between optical and $\gamma$-ray fluxes in 2008--2010 \citep{hay12}.  
The lack of overall correlation between the optical light 
curve and the radio (mm-band), X-ray, and $\gamma$-ray light curves suggests that
they are produced by different populations of electrons, and most likely at different locations.
Moreover, the apparent 
linear relation between SEDs for Periods A and C is merely a coincidence. 
As the overall radiative output from 3C~279 is always dominated by the $\gamma$-ray 
emission, the systematic long-term increase in the optical 
flux could be due to a systematic increase of the magnetic field strength, 
or a systematic decrease of the external radiation energy density. 
The former option would be problematic if the jet magnetization remains constant, 
which would lead to an increase in the total jet power, and ultimately to 
an increase in the $\gamma$-ray luminosity, which is not observed.

\section{Conclusions}
\label{sec_conc}

We report the results of observations of the well-studied $\gamma$-ray 
luminous blazar 3C~279 at the end of 2013 
and beginning of 2014, when the object entered a bright and active state.  
The \Fermi-LAT observations revealed multiple distinct, bright 
flares, and recorded the highest $\gamma$-ray flux state of 
the source since the launch of \Fermi, at $F(E>100 \rm MeV)$ reaching 
$10^{-5}$ photons\,cm$^{-2}$\,s$^{-1}$ on 2013 December 20 and 2014 April 03.
The high flux of 
the source allowed us to integrate the $\gamma$-ray flux on time scales
as short as one \Fermi-LAT orbit (96 min).
This in turn allowed us to 
establish the variability time scales to be as short as 
$t_{\rm var} \simeq 2$ hours.
One of these flares revealed an unprecedentedly
hard $\gamma$-ray spectrum, with a photon index $\Gamma_\gamma \simeq 1.7$, unusual for this kind of source.

Two \NuSTAR\ observations provided the first precise measurement of the hard X-ray spectrum of 3C~279 up to 70 keV. 
The \NuSTAR\ observations were complemented by more frequent \Swift\ observations.
The best-fit model for the joint spectra by \NuSTAR\ and \Swift-XRT during the first \NuSTAR\ observation (2013 December 16) was consistent with a simple power law, which usually has been observed in past X-ray observations involving {\it Suzaku} and \XMM~\citep{hay12}.
On the other hand, the second simultaneous \NuSTAR\ and \Swift\ observations at the end of 2013
revealed a spectral structure that was harder ($\Gamma_{\rm X1} \simeq 1.37$) below $\simeq 3.5\;{\rm keV}$ 
and softer ($\Gamma_{\rm X2} \simeq 1.76$) above that energy. 
Such a spectral structure has not been observed in 3C~279 in any previous X-ray observations.
In addition, the second \NuSTAR\ observation (2013 December 31) indicated an increase of the X-ray flux by $\sim 50$\% during the 1-day pointing. In the soft X-ray data from \Swift-XRT, we found a clear correlation between the flux and photon index with a harder-when-brighter trend.
More detailed studies offer potential for better understanding of the origin of X-ray emission in FSRQ blazars.

The optical flux of the source steadily increased since the beginning of this multiwavelength campaign, 
but did not show clear, large-amplitude
flares such as those seen in $\gamma$-rays.  
The optical flux of the source does not appear clearly correlated 
with the $\gamma$-ray flux, in contrast to the behavior 
measured in 2008--2010~\citep{hay12}.
It is possible that the optical flux might be arising in 
multiple locations along the jet, as expected if the electrons radiating in the optical are of different energy than those making the X- and $\gamma$-rays, which is suggested by the apparent lack of correlation between the optical flux and the optical polarization degree.

We modeled the broad-band spectrum of the source at multiple epochs, 
including two very bright $\gamma$-ray flares, and the two \NuSTAR\ pointings.
The very hard $\gamma$-ray spectrum and very high Compton dominance during 
the first $\gamma$-ray flare are very challenging to explain in the standard one-zone 
synchrotron plus SSC/ERC model, requiring electron energy distribution 
index $p \simeq 1$, a high jet power compared to the accretion disk luminosity $L_{\rm j}/L_{\rm d} \gtrsim 10$, and a very low magnetic fraction of the jet 
power $L_{\rm B}/L_{\rm j} \lesssim 10^{-4}$.
In addition, no single-zone 
modeling of any single epoch of the broad-band SED can satisfactorily 
explain the production of X-ray emission at the same time as it explains the optical 
and $\gamma$-ray emission.  This conclusion is consistent with the finding that variations in the X-ray flux of 3C~279 are not always simultaneous with variations in the $\gamma$-ray or optical fluxes. 

\acknowledgements
The \textit{Fermi} LAT Collaboration acknowledges generous ongoing support
from a number of agencies and institutes that have supported both the
development and the operation of the LAT as well as scientific data analysis.
These include the National Aeronautics and Space Administration and the
Department of Energy in the United States, the Commissariat \`a l'Energie Atomique
and the Centre National de la Recherche Scientifique / Institut National de Physique
Nucl\'eaire et de Physique des Particules in France, the Agenzia Spaziale Italiana
and the Istituto Nazionale di Fisica Nucleare in Italy, the Ministry of Education,
Culture, Sports, Science and Technology (MEXT), High Energy Accelerator Research
Organization (KEK) and Japan Aerospace Exploration Agency (JAXA) in Japan, and
the K.~A.~Wallenberg Foundation, the Swedish Research Council and the
Swedish National Space Board in Sweden.
Additional support for science analysis during the operations phase is gratefully acknowledged from the Istituto Nazionale di Astrofisica in Italy and the Centre National d'\'Etudes Spatiales in France.

This work was partially supported under the NASA contract no.\ NNG08FD60C, and made use of observations from the \NuSTAR\ mission, a project led by California Institute of Technology, managed by the Jet Propulsion Laboratory, and funded by NASA.  We thank the \NuSTAR\ Operations, Software and Calibration teams for support of the execution and analysis of these observations. This research has made use of the \NuSTAR\ Data Analysis Software (NuSTARDAS) jointly developed by the ASI Science Data Center (ASDC, Italy) and the California Institute of Technology (USA).
This research has made use of the XRT Data Analysis Software (XRTDAS) developed under the responsibility of the ASI Science Data Center (ASDC), Italy.
The Submillimeter Array is a joint project between the Smithsonian Astrophysical Observatory and the Academia Sinica Institute of Astronomy and Astrophysics and is funded by the Smithsonian Institution and the Academia Sinica.

K.N.\ was supported by NASA through Einstein Postdoctoral Fellowship grant number PF3-140130 awarded by the {\it Chandra} X-ray Center, which is operated by the Smithsonian Astrophysical Observatory for NASA under contract NAS8-03060. 
M.B.\ acknowledges support from NASA Headquarters under the NASA Earth and Space Science Fellowship Program, grant NNX14AQ07H.

Facilities: \facility{Fermi-LAT}, \facility{NuSTAR}, \facility{Swift}, \facility{SMARTS}, \facility{Kanata}, \facility{SMA}

\end{document}